\documentclass[useAMS,usenatbib,usegraphicx]{mn2e}

\usepackage{natbib}
\usepackage{graphicx}
\usepackage[latin1]{inputenc}
\usepackage{color}
\usepackage{times}
\usepackage{setspace}
\newif\ifAMStwofonts
\AMStwofontstrue
\definecolor{red}{rgb}{1,0.,0.}

\newcommand{\mor}{{\sc morgana}}
\newcommand{\gs}{{\sc grasil}}
\newcommand{\msun}{{\rm M}_\odot}

\def\lesssim{\lower.5ex\hbox{$\; \buildrel < \over \sim \;$}}
\def\gtrsim{\lower.5ex\hbox{$\; \buildrel > \over \sim \;$}}

\title[EROs in \mor] {The active and passive populations of Extremely
  Red Objects.}

\author[Fontanot \& Monaco]{
  \parbox[t]{\textwidth}{ 
    Fabio Fontanot$^{1,2}$ \& Pierluigi Monaco$^{1,3}$
    }
    \vspace*{6pt}\\
    $^1$ INAF-Osservatorio Astronomico, Via Tiepolo 11, I-34143 Trieste, Italy \\
    $^2$ MPIA Max-Planck-Institute f\"ur Astronomie, Koenigstuhl 17, 69117 Heidelberg, Germany\\
    $^3$ Dipartimento di Fisica, Universit\`a di Trieste, via Tiepolo 11, 34143 Trieste, Italy \\
    email: fontanot@oats.inaf.it, monaco@oats.inaf.it}

\begin{document}
\date{Accepted ... Received ...}

\maketitle

\begin{abstract} 
  The properties of galaxies with the reddest observed $R-K$ colors
  (Extremely Red Objects, EROs), including their apparent division
  into passive and obscured active objects with roughly similar number
  densities, are a known challenge for models of galaxy formation.  In
  this paper we produce mock catalogues generated by interfacing the
  predictions of the semi-analytical {\mor} model for the evolution of
  galaxies in a $\Lambda$CDM cosmology with the spectro-photometric +
  radiative transfer code {\gs} and Infrared (IR) template library to
  show that the model correctly reproduces number counts, redshift
  distributions and active fractions of $R-K>5$ sources. We test the
  robustness of our results against different dust attenuations and,
  most importantly, against the inclusion of TP-AGB stars in Simple
  Stellar Populations (SSPs) used to generate galaxy spectra, and find
  that the inclusion of TP-AGBs has a relevant effect, in that it
  allows to increase by a large factor the number of very red active
  objects at all color cuts. We find that though the most passive and
  the most obscured active galaxies have a higher probability of being
  selected as EROs, many EROs have intermediate properties and the
  population does not show bimodality in specific star formation rate
  (SSFR).  We predict that deep observations in the Far-IR, from 100
  to 500 $\mu$m, are the most efficient way to constrain the SSFR of
  these objects; we give predictions for future {\it Herschel}
  observations, and show that a few objects will be detected in deep
  fields at best.  Finally, we test whether a simple evolutionary
  sequence for the formation of $z=0$ massive galaxies, going through
  a sub-mm-bright phase and then a ERO phase, are typical in this
  galaxy formation model. We find that this sequence holds for
  $\sim25$ per cent of $z = 0$ massive galaxies, while the model
  typically shows a more complex connection between sub-mm, ERO and
  massive galaxies.
\end{abstract}


\section{Introduction}\label{intro}
The wealth of data coming from recent multiwavelength surveys has made
it possible to study in detail distant galaxy populations from their
UV-optical-Infrared broad-band spectral energy distributions (SED).
Among the various categories that have been observationally defined by
means of color selection techniques, the Extremely Red Objects (EROs)
have attracted wide interest.  Defined on the basis of of their very
red $R-K$ color (in the Vega system $R-K>5$ at least), they were first
addressed in the context of deep near-infrared surveys
\citep{Elston88, McCarthy92}. Such red colors are expected to arise as
a combination of an intrinsically red SED and a large K-correction at
$1<z<2$. Then, EROS were regarded as natural counterparts of local
elliptical galaxies \citep[e.g.][]{Cimatti02a}, Moreover, their
relatively large $K$-band fluxes suggested high stellar masses,
comparable with local ellipticals, leading to the interpretation of
EROs as already assembled elliptical galaxies and the redshift range
$z>2$ as their formation epoch. The correct prediction of such an
early and efficient formation of massive galaxies at high redshift is
a well known challenge for models of galaxy formation based on the CDM
cosmology: in fact early implementations predicted lower space
densities of EROs than observed \citep[e.g.,][]{Firth02,Smith02}.

On the other hand, it soon became evident that the same red $R-K$
colors might be linked to a completely different class of objects such
as the active (starburst) galaxies embedded in dust shells that cause
high extinction of the SEDs
\citep{Cimatti02a,PozzettiMannucci00,Smail02}.  Interestingly enough,
the relative contribution to the total space density by the two
sub-populations is similar, with a passive fraction of $48$ per cent
at $K<19.2$ (\citealt{Kong09}, but a ratio between 30 and 70 per cent
has been reported also by \citealt{Cimatti02a} and \citealt{Smail02}),
and $58$ per cent at $K<20.3$ \citep{Miyazaki03}.  However, splitting
the observed ERO population in two is a difficult task.  The cleanest
way relies on high signal-to-noise spectroscopy, able to reveal either
emission lines linked to ongoing star formation or the spectral breaks
typical of aged stellar populations (see e.g., \citealt{Cimatti02a}).
Alternatively, the presence of a starburst may be revealed by
additional imaging in the far-infrared, where the absorbed light is
reprocessed \citep{Kong09}, or indirectly through the morphology of
these objects \citep{StiavelliTreu01,Cimatti03}. Presently, all these
techniques can be successfully applied only to a small fraction of the
overall population, typically limited to the brightest sources.

As mentioned above, the redshift distribution and the expected
physical properties of the two ERO sub-populations have been a
long-standing problem for theoretical models of galaxy formation
\citep{Cimatti02a,Somerville04b}.  More recent models, based on
improved treatment of baryonic physics
\citep{DeLucia06,Bower06,Menci06,Monaco07,Lagos08}, are able to give a
better representation of the formation and assembly of massive
galaxies at $z<3$. Only a few of these models have been directly
tested by comparing the predicted properties of EROs with data: in
most cases a reasonable success in reproducing the abundance of
passive EROs has been claimed, giving support to the interpretation of
these objects as passive massive galaxies at $z\sim1-2$ that are the
progenitors of local elliptical galaxy. At the same time, the same
models do not produce a sufficient number of active EROs.  More
specifically, \citet{Nagamine05} implemented sub-grid baryonic physics
in both a Lagrangian and Eulerian code; they found the correct
abundance of massive galaxies at $z\sim 2$, but could find almost no
passive ERO, and in order to produce enough red galaxies they were
forced to assume a dust attenuation as high as an equivalent
$E(B-V)=0.4$ for the whole mock catalogue. \citet{Kang06} correctly
reproduced the space density of massive galaxies at $z\sim 2$, but
underpredicted the number of very red galaxies, in particular of the
active class. \citet{Menci06} proposed a more successful model able to
reproduce the color distribution of galaxies at $z<1$ and the space
density of massive galaxies at $1<z<2$ at the same time.  This model
is able to reproduce the ERO number counts and redshift distribution,
but it predicts a relatively high fraction of passive objects
($\sim$80 per cent at $K<22$).  More recently, \citet{GonzalezPerez09}
compared different implementations of the {\sc galform} model
\citep{Cole00} to observational data of the ERO population.  Their
best fit model \citep[i.e.,][]{Bower06} produced the correct number of
EROs, even when a very red cut (up to $R-K>7$) was used, but again the
predicted ERO population was dominated by passive objects, with only a
small fraction of active galaxies.  On the other hand the model by
\citet{Baugh06}, which gives a good fit of the sub-mm galaxy
population, was found to significantly underestimate ERO number
counts.

A remarkable attempt to put quasars, bright sub-mm galaxies, EROs, and
low-$z$ ellipticals into a unified scenario was proposed by
\citet{Granato04}. In their model a typical low-z elliptical forms the
bulk of its mass in an intense obscured starburst phase at $z\sim2$,
when it is visible in the sub-mm band.  In the galaxy nucleus, star
formation triggers accretion onto a seed black hole; the end of the
star formation phase is caused by feedback from the resulting quasar,
able to wipe away the interstellar medium from the galaxy and quench
star formation. After $\sim1$ Gyr the galaxy gets very red and becomes
an ERO. It then evolves passively into a typical local elliptical
galaxy.

The above mentioned papers have clarified the role of the ERO
population as a powerful test for theoretical models of galaxy
formation. These should reproduce both their global number and the
diversity of their physical properties.  A successful model would be
helpful in investigating the relation of EROs with other
observationally selected galaxy populations, in particular with the
sub-mm galaxies.  Apparently, the most challenging requirement for
published model is to produce a high fraction of active EROs,
especially when deep samples are concerned.  In principle, active EROs
should be easy to produce in a hierarchical model of galaxy formation,
because the redshift range where we observe them corresponds to the
peak of the cosmic star formation rate density. \citet{Kong06} and
\citet{Nagamine05} interpret the tension between their models and data
as a sign of over-simplified recipes in treating baryonic physics or
dust attenuations. Regarding the latter point, \citet{Fontanot09a}
showed that currently used recipes for dust attenuation, calibrated
using low-$z$ samples, provide very scattered extinction values at
$z\sim 2$ and, more interestingly, do not match the predictions of the
{\gs} code \citep{Silva98} that explicitly solves the equations of
radiative transfer. Another important point is the contribution of the
TP-AGB stars to the synthetic SEDs of model galaxies (see
e.g. \citealt{Maraston05}), which is not included in the above
mentioned models. It has been shown by \citet{Tonini08, Tonini09} that
the inclusion of this extreme phase of stellar evolution has a strong
impact on the expected photometry and colors of simulated galaxies at
$1 \la z \la 4$, where intermediate-age stellar populations dominate
the restframe $K$-band starlight. In the case of EROs, an increase of
their number density is expected when TP-AGBs are included due to the
reddening of rest-frame $V-K$. All these elements critically affect
the modeling of galactic SEDs: they are therefore a primary concern
for the interpretation of the physical ingredients of models.

In this paper we use the {\mor} model (\citealt{Monaco07}, as updated
in \citealt{LoFaro09}), interfaced with the {\gs} spectro-photometric
+ radiative transfer code \citet{Silva98} and to Infrared (IR)
template library \citep{Rieke09}, to produce mock samples of EROs. We
both use the {\it Padova} \citep{Bertelli94} and the \citet[][M05
  hereafter]{Maraston05} library of SSPs. We find reasonable agreement
with data in terms of number counts, redshift distribution and
fraction of active objects, especially when M05 library is used
(details on the effect of this modification on the whole galaxy
population will be the subject of a forthcoming paper). This prompts
us to use the model to investigate the physical properties of these
object.  We address whether there is a true active/passive bimodality
of the ERO population and what is the connection between EROs, sub-mm
galaxies and massive galaxies at $z=0$. We finally give predictions of
ERO properties for future surveys with the {\it Herschel} satellite.

This paper is organized as follows. In sec.~\ref{sec:model} we recall
the main feature of the theoretical models {\mor} and {\gs}. In
sec.~\ref{sec:results} we present our results: in
sec.~\ref{sec:ero_prop} we analyze the properties of the model ERO
sample and in sec.~\ref{sec:sequence} we discuss the connection
between the ERO population, the sub-mm bright galaxies and the massive
galaxies at $z=0$.  Finally in sec.~\ref{sec:final} we give our
conclusions.  Throughout the paper we assume magnitudes are in the
Vega system (unless otherwise stated), and a cosmological model
consistent with WMAP3 results ($\Omega_0=0.24$, $\Omega_\Lambda=0.76$,
$h=0.72$, $\sigma_8=0.8$, $n_{\rm sp}=0.96$, \citealt{Komatsu09}).

\section{Model}\label{sec:model}

We use the semi-analytical MOdel for the Rise of GAlaxies aNd AGNs
(\mor), originally presented in \citet{Monaco07}, and updated in
\citet{LoFaro09}, where it was optimized for the WMAP3 $\Lambda$CDM
cosmology used in this paper, {\it Padova} SSP library and a
\citet{Chabrier03} stellar Initial Mass Function (IMF). All details
are reported in the former paper, while a brief outline is given in
the latter; we refer to them for a description of the model.  The only
new extension we present in this paper is the use of M05 library as an
alternative to the {\it Padova} ones.  The predictions of {\mor} were
compared with a number of observations in a series of papers
\citep{Fontanot06, Fontanot07b, Fontanot09b}.  In those papers the
model was calibrated to reproduce the local stellar mass function and
the star formation rate density as a function of redshift, while some
parameters (especially those regulating accretion onto central black
holes) were fixed by requiring a good fit of AGN luminosity functions.
Among the most interesting successes of the model, {\mor} is able to
correctly reproduce, with the same combination of parameters, (i) the
evolution of the AGN population and in particular the optical and
X-ray luminosity functions; (ii) the redshift distribution and
luminosity functions of $K$-selected samples in the redshift range
$0<z<3$; (iii) the number counts of $850 \mu m$-selected sources
assuming a conservative Salpeter or Chabrier IMF.  Moreover, the
progressive build-up of bright Lyman-break galaxies at $z\ga4$ is
recovered after the optimization of dust attenuation parameters that
have a dramatic influence on rest-frame UV radiation.

$K$-band and sub-mm counts of {\mor} galaxies were shown by
\cite{Fontanot07b} for the original version of the model. With the
change of cosmology, IMF and the improvements given in
\cite{LoFaro09}, $K$-band number counts are very similar to those
presented in \cite{Fontanot09b}, with a better fit at $K\sim18$,
roughly corresponding to the knee of the local luminosity function.
The agreement of the redshift distribution of $K<20$ galaxies with
observations shows a slight but appreciable improvement.  Though the
stellar mass function is well recovered at high masses with the
improved model \citep{Fontanot09b}, we still produce an excess of
bright galaxies in the $K$-band local luminosity function (for
$K<-25.5$, amounting to $\sim 1$ dex at $M_K = -26.5)$. This is due to
the too high specific star formation rate of local massive galaxies
(see below), which results in mass-to-light ratios $M_\star/L_K \sim
0.5 M_\odot/L_\odot$, to be compared, e.g., with the value of $0.75$
assumed by \citet{Cole01} to convert the local $K$-band luminosity
function into a stellar mass function (see \citealt{Fontanot07b} for a
complete discussion of this issue).  Sub-mm counts are depressed by a
factor of 0.5 dex when passing from Salpeter to Chabrier IMF, while
their redshift distribution does not change appreciably: this implies
that in the new model sub-mm number counts are consistent with data up
to a flux of $3$ mJy, while they are undepredicted at brighter fluxes.
  
Previous papers have found a number of points of severe tension
between model and data, some of which are also present for most
similar models.
\begin{enumerate}
\item{The colors of model massive galaxies at low redshift are too
  blue \citep{Kimm08}: this is a peculiar problem of {\mor} and is
  mainly due to the modeling of AGN feedback: because star formation
  is required to trigger black hole accretion, an efficient AGN
  feedback requires some small amount of star formation to take place,
  and this makes massive galaxies bluer than observed.}
\item{The color distribution of low-z satellite galaxies is not
  correctly reproduced, with the fraction of red satellites
  outnumbering the corresponding blue fraction \citep{Weinmann06a}.
  This is directly related to ``strangulation'', i.e. the assumption
  that satellites are stripped of all their hot gas when they infall
  on a larger group. \citet{Kimm08} recently showed that this problem
  is common to most semi-analytic models of galaxy formation.}
\item{The so-called ``downsizing'' trend in galaxy formation
  (e.g. more massive galaxies forming on a shorter timescale and at
  higher redshift with respect to lower mass counterparts) is not
  fully recovered. The tension between model predictions and
  observations is significant for low-to-intermediate stellar mass
  galaxies ($M_\ast < 10^{11}M_\odot$): they form too efficiently at
  high redshift \citep[see also]{LoFaro09} and are too passive at
  later time, so that the model overpredicts their observed stellar
  mass function at $z>1$.  As a result such small galaxies are
  predicted to host too old stellar population at $z=0$.
  \citet{Fontanot09b} showed that this behavior is common to most
  published models based on $\Lambda$CDM (i.e. \citealt{Wang08,
    Somerville08}).}
\item{The number of massive galaxies at $z>3$ is underpredicted by
  {\mor}: despite the evolution of the $K$-band luminosity function is
  reasonably reproduced by the model up to $z \sim 2$
  \citep{Fontanot07b}, \citet{Cirasuolo07,Cirasuolo08} showed that at
  higher redshifts the bright end is systematically underpredicted. A
  similar effect is seen in the evolution of the stellar mass function
  \citep{Fontanot09b}. This tension is expected to weaken when TP-AGB
  stars are included in the population synthesis model used to
  generate the SEDs \citep{Tonini09}, though \cite{Marchesini09} find
  that this is not sufficient to remove the discrepancy.}
\item{Specific star formation rates of active galaxies at $z\sim2$ in
  {\mor} galaxies are higher than those of other models, but they may
  still be a factor of $\sim3$ lower than observed ones
  \citep{Santini09,Fontanot09b}, though this difference may well be due
  to systematics in star formation rate estimators. }
\item{Higher specific star formation rates are clearly the key for the
  success of {\mor} in reproducing the elusive sub-mm counts, but the
  redshift distribution of the most luminous sub-mm sources peaks at
  lower redshift than observed by \citet{Chapman05}.  This discrepancy
  is clearly connected with the lack of downsizing in the model:
  massive starbursts should take place earlier than average, not later
  as in our model.}
\end{enumerate}

All these issues can influence our results on the ERO population; for
instance, a population of strangulated satellites could artificially
boost the number of EROs, while residual star formation in massive
galaxies could depress it. We will keep in mind all these
discrepancies when interpreting our results.

\begin{figure*}
  \centerline{
    \includegraphics[width=18cm]{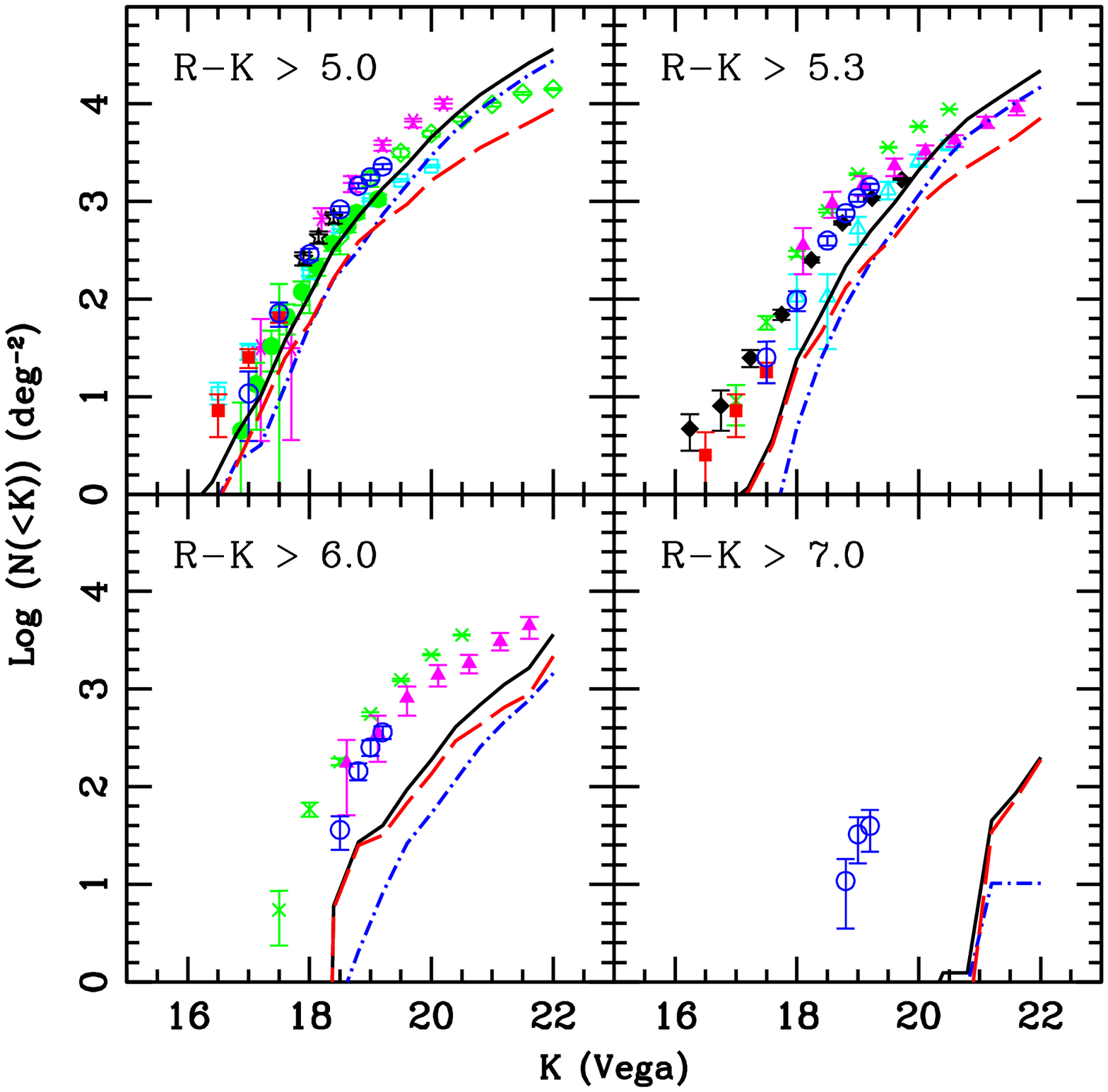} }
  \caption{ERO cumulative number counts in the reference model for
    four $R-K$ (Vega) color cuts, as given in the panels.
    Observational datapoints are taken from \citet[blue empty
      circles]{Daddi00}, \citet[cyan empty triangles]{Smail02},
    \citet[magenta filled triangles]{Smith02}, \citet[black empty
      diamonds]{Roche03}, \citet[magenta asterisks]{Miyazaki03},
    \citet[red filled squares]{Vaisanen04}, \citet[black
      stars]{Brown05}, \citet[green crosses]{Simpson06}, \citet[green
      filled circles]{Kong06}, \citet[cyan empty squares]{Lawrence07},
    \citet[black filled diamonds]{Conselice08}. Solid black lines
    refer to the prediction for the whole ERO population, red dashed
    and blue dot-dashed lines refer to the passive and active classes
    respectively.}
  \label{fig:nceros}
\end{figure*}
\begin{figure*}
  \centerline{
    \includegraphics[width=18cm]{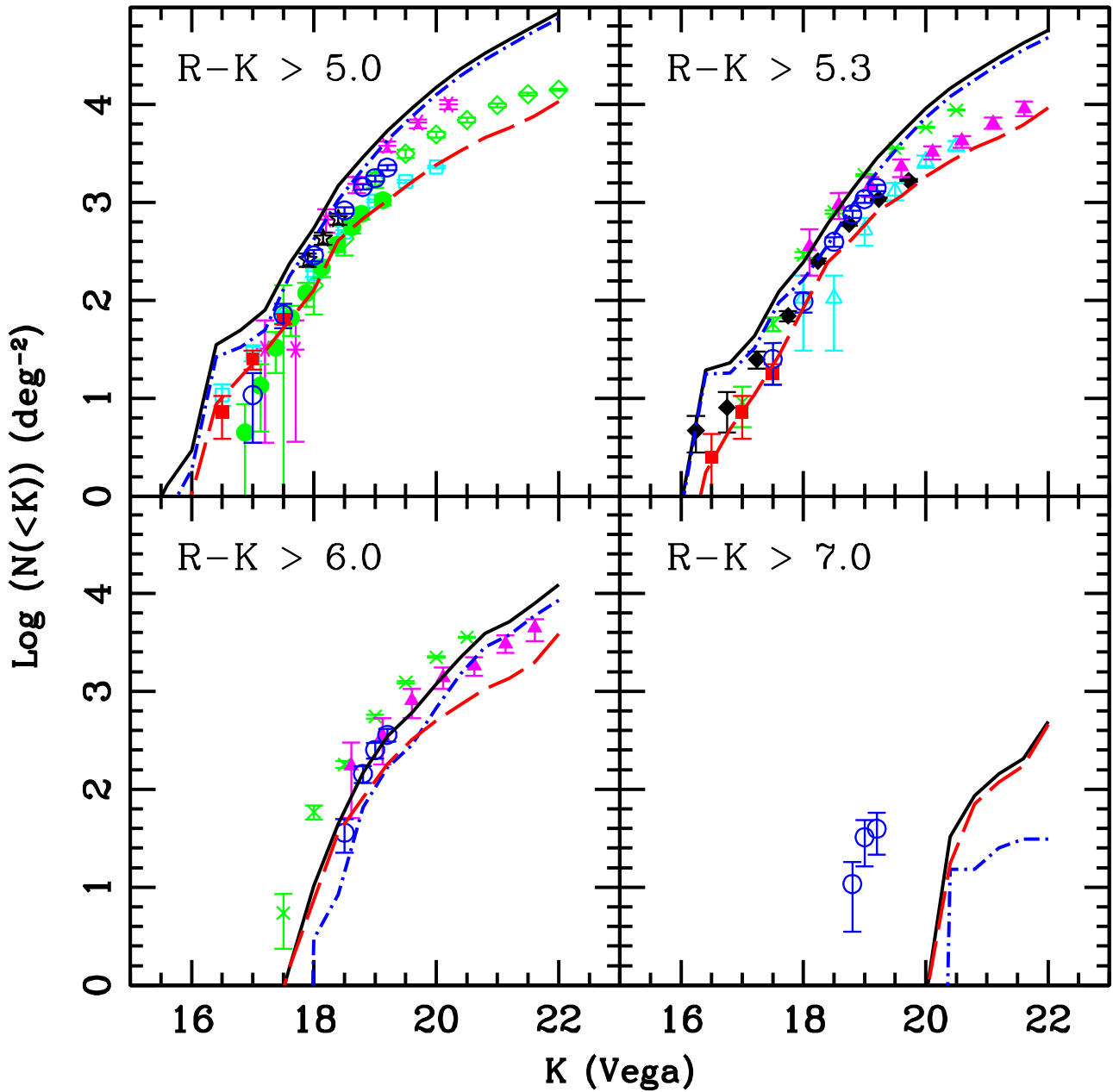} }
  \caption{ERO cumulative number counts in the model using M05 SSP
    library for four $R-K$ (Vega) color cuts. Symbols and lines are as
    in figure~\ref{fig:nceros}.}
  \label{fig:nceros_mar}
\end{figure*}

\subsection{Generation of mock catalogues}\label{sec:mock}
\begin{figure*}
  \centerline{ \includegraphics[width=9cm]{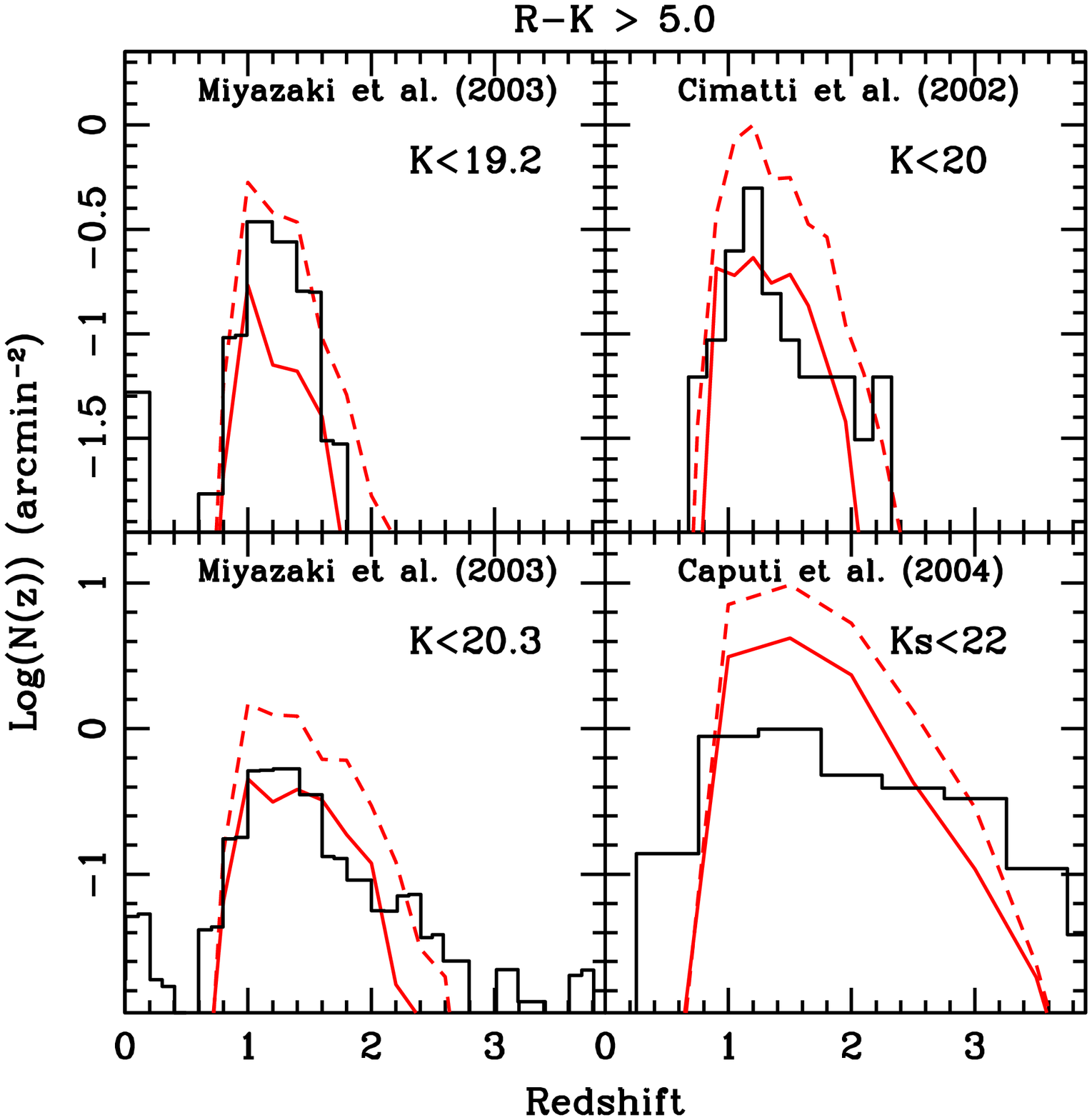}
    \includegraphics[width=9cm]{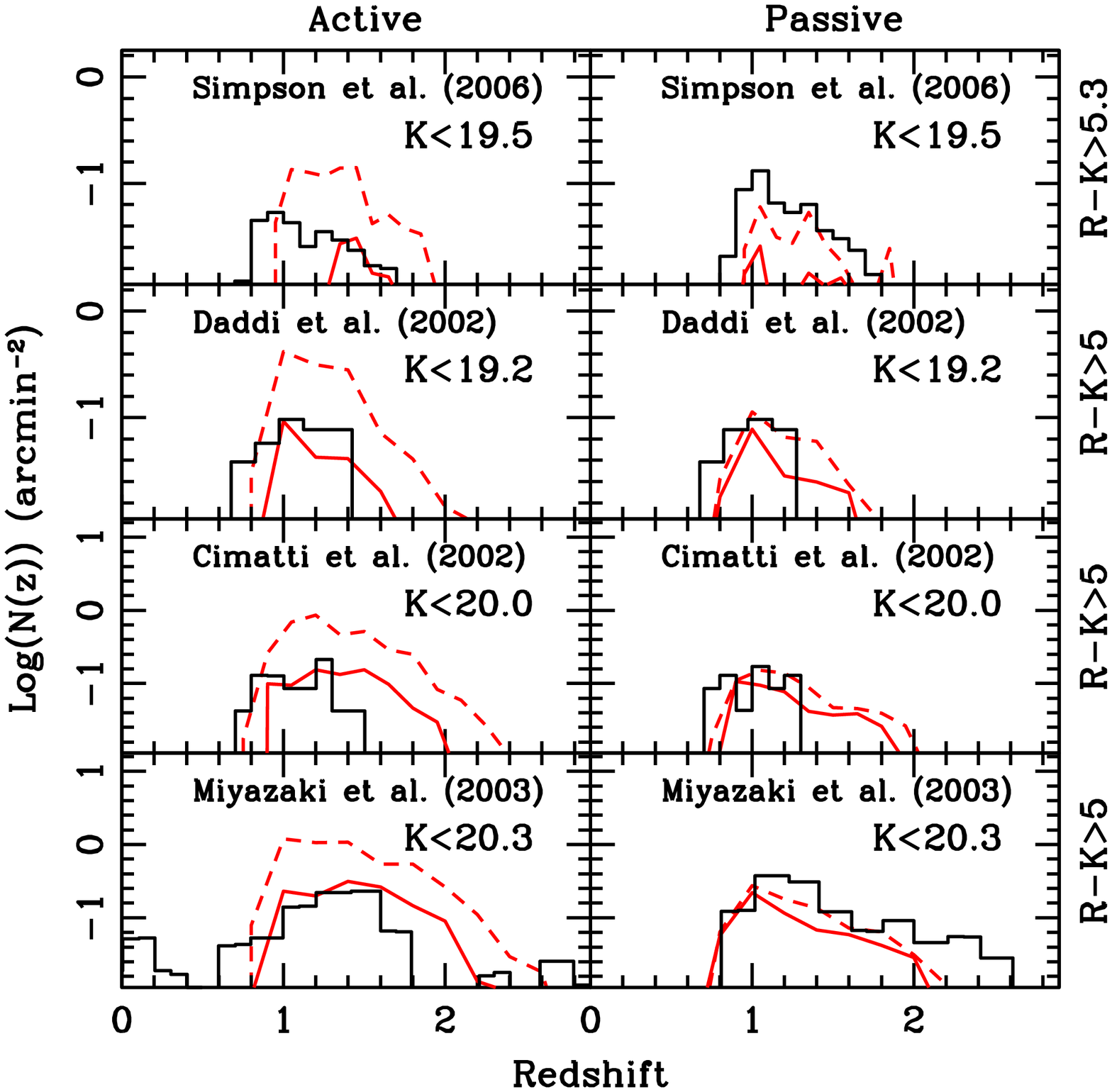} }
  \caption{Left: predicted and observed ERO redshift distributions at
    the flux limits given in the panels. Data are from \citet[
      spectroscopic completeness 67\%]{Cimatti02c}, \citet[
      photometric redshifts]{Miyazaki03}, \citet[photometric
      redshifts]{Caputi04}. The continuous and dashed lines refer to
    the model predictions with {\it Padova} and M05 SSP library,
    normalized to the same area of the corresponding surveys. Right:
    observed and predicted ERO redshift distributions splitted into
    active and passive populations. Data taken from
    \citet[spectroscopic redshifts]{Daddi02}, \citet{Cimatti02c},
    \citet{Miyazaki03} and \citet[photometric redshifts]{Simpson06}.}
  \label{fig:zd_eros}
\end{figure*}

We generate our main mock catalogue using a {\sc pinocchio}
\citep{Monaco02} simulation of a $200 Mpc$ comoving box, sampled with
$1000^3$ particles.  The particle mass is then $\sim 10^9 \msun$ and
the smallest halo we consider contains 50 particles, for a mass of
$5.1 \times 10^{10} \msun$.  The typical stellar mass of the central
galaxy contained in the smallest DM halo at $z=0$ is $\sim 5 \times
10^8 \msun$.  The same sampling strategy to generate deep fields as in
\citet{Fontanot07b} is adopted here, with the only difference that, in
order to have a sufficient number of EROs, all merger trees in the box
are used ($w_{\rm tree}=1$ in the original notation, see the original
paper for more details on sampling strategy and catalogue
generation). The final reference mock catalogue consists in $\sim
240000$ model galaxies, covering a redshift range $0<z<8$.

{\mor} provides predictions on galaxies in terms of physical, non
observable quantities.  We then interface {\mor} with the
spectro-photometric + radiative transfer code {\gs} \citep{Silva98}.
This code computes the synthetic UV-to-radio SEDs of galaxies from the
known star formation history, cold gas mass, cold gas metal mass and
size, separately provided for a bulge and a disc component.  Further
assumptions are made on bulge profile, vertical width of the disc,
dust distribution in molecular clouds and diffuse gas.  Stars are
assumed to be born within the optically thick molecular clouds and to
gradually escape from them as they get older.  This gives rise to an
age-selective extinction, with the youngest and most luminous stars
suffering larger dust extinction than older ones. The radiative
transfer of starlight through dust is computed assuming an
axisymmetric bulge-disc system, yielding the emerging SED.  Some of
the parameters needed by {\gs} are not directly provided by {\mor}.
We fix their values as in \citet{Fontanot07b}, where a choice was made
which is good for the average galaxy population: in particular we use
an escape time-scale of young stars for the parent molecular clouds
equal to $10^7$ yr and a $0.5$ fraction of cold gas mass in the form
of molecular clouds. These values differ from the choice done in
\citet{LoFaro09}, which was optimized for high-$z$ Lyman-break
galaxies: escape time equal to $0.3 \times 10^7$ yr and a $0.95$
molecular fraction.  As shown by \citet{Silva05}, the final fraction
of active EROs in their model depends strongly on these unconstrained
parameters: this is especially true for the objects at $z>1$, which
require a detailed computation of the highly uncertain dust
obscuration in the restframe UV. They reported that obscuration is
especially sensitive to the assumed molecular fraction. Because our
aim in this paper is to test the predictions of already published
versions of the model and not to find the best-fit parameters for this
specific observables, we decided to use the parameters of
\citet{Fontanot07b}, and test how results change using the high
molecular fraction of \citet{LoFaro09}.  Apparent magnitudes are
computed by adding a fiducial 0.1 mag random Gaussian error.  Number
counts and redshift distributions are then constructed as in
\citet{Fontanot07b}. It is worth stressing that our reference
      {\mor}+{\gs} model is calibrated using the {\it Padova} SSP
      library. Here, we also produce a catalogue using the M05 SSP
      library (which adopts a \citet{Kroupa93} IMF - we neglect the
      small differences with the Chabrier IMF), to check the effect of
      the inclusion of the TP-AGBs on the ERO statistics. 

Given the low space density of the ERO population, a higher sampling
of model galaxies is required with respect to the mock catalogues
defined in \citet{Fontanot07b}. In order to speed up computations (see
\citealt{Fontanot07b} for a general discussion about the time
requirements of the {\mor}+{\gs} algorithm), we modify our standard
approach as follows.  For each model galaxy we compute the
UV-to-Near-IR SED with {\gs}, including dust attenuation but excluding
the computation of the Far-IR emission due to heated dust, which is by
far the most time-consuming step. With this choice we are then able to
generate the synthetic SEDs in a few seconds instead of $\sim5-10$
min. We then compute the total absorbed starlight and we use it to
select an appropriate IR template from the \citet{Rieke09} library.
We renormalize the chosen template to the total absorbed starlight and
use it to model IR emission, which is significant for fluxes at
$\lambda > 3 \mu$m. As shown in \citet{Fontanot09c}, this procedure
provides an estimate for the SED which is however, for high-redshift
galaxies, biased low by $\sim0.2$ dex at high sub-mm luminosities but
is correct on average at fluxes of $\la 1$ mJy, with respect to {\gs}
code. As a cross check, we apply a full {\gs} calculation to a much
smaller mock catalogue of $\sim32000$ galaxies and compare the
resulting number counts and redshift distribution with those based on
IR templates. No significant differences were found at
$\lambda<850\mu$m. The normalization of sub-mm number counts is
correctly reproduced at fluxes $f_{850 \mu m} \sim 0.5$ mJy: at
brighter fluxes the counts are underpredicted by a factor of 4 at
$f_{850 \mu m} \sim 3$ mJy, due to the low biasing of this procedure
with respect to the predictions of the full {\gs} calculations, while
at fainter fluxes ($> 0.1$ mJy) they are overpredicted by a similar
amount.

\section{Results}
\label{sec:results}

In fig.~\ref{fig:nceros} we compare the predicted ERO cumulative
number counts (black line) with available data \citep{Daddi00,
  Smith02, Smail02, Roche03, Vaisanen04, Brown05, Kong06, Simpson06,
  Lawrence07, Conselice08}, for color cuts $(R-K)>5.0$, $5.3$, $6.0$
and $7.0$. We first consider the predictions of our reference model:
for the first color cut, the agreement is good down to $K\sim20$,
where the model starts to overpredict number counts.  This behavior is
very similar to that of the total sample (figure 2 of
\citealt{Fontanot07b}). We stress that this level of agreement is
obtained, with no parameter tuning, from the same model used in
\cite{Fontanot09b} and \cite{LoFaro09}.  The agreement worsen at
redder colors: the brightest sources at $(R-K)>5.3$ are underpredicted
by an order of magnitude, lowering to a factor of 3 at $K\sim18$; we
find an overall underestimate by an order of magnitude for $R-K>6$ and
a much more severe one for $R-K>7$.

We then split the whole ERO population into active and passive
subsamples using the instantaneous specific star formation, ${\rm
  SSFR} = {\rm SFR} / M_\star$.  As a threshold between the two
populations we assume the same criterion as in \citet[${\rm
    SSFR}=10^{-11} yr^{-1}$]{Brinchmann04} at all redshifts: at the
depth of $K=22$ and for $R-K>5.0$ we classify $\sim 68$ and $\sim32$
per cent of model EROs as active and passive respectively. At $K<20.3$
the relative fractions of active and passive EROs become $\sim 62$ and
$\sim38$ per cent to be compared with the results of \citet[][, $42$
  and $58$ per cent respectively]{Miyazaki03}, while at $K<19.2$ the
fractions are $\sim 53$ and $\sim 47$, consistent with the results of
\citet[][both $50 \pm 20$ per cent]{Cimatti02a}. Finally at $R-K>5.3$
and $K<20.5$ our model predicts a relative fraction of $56\%$ active
EROs, to be compared with the results of \citet{Smail02}, who found an
active fraction ranging from a lower limits of $30\%$ up to an upper
limit of $60\%$.  Figure~\ref{fig:nceros} gives the predicted
cumulative number counts for the two classes of EROs; for the first
color cut passive EROs (red dashed line in the figure) are more
numerous at bright fluxes, active ones (blue dot-dashed line) start to
dominate at $K>18.5$.

To test the stability of our results against different SSP libraries
and models of dust attenuation we repeat the analysis using different
assumptions for these ingredients. We create three mock catalogues by
interfacing the same {\mor} realization with the following
modifications to our standard {\gs} setup:
\begin{enumerate}
\item{we consider a higher molecular fraction (0.95) and a lower
  escape time for stars (3 Myr) in active galaxies, as used by
  \citet{LoFaro09} to reproduce the properties of bright Lyman Break
  Galaxies;}
\item{we model dust extinction using the prescription proposed by
  \citet{DeLucia07b}, which combines their fitting formula for the
  optical depth in the $V$-band with a slab geometry and a composite
  age-dependent attenuation law (i.e. assuming a Milky Way extinction
  curve for the diffuse ``cirrus'' dust and the
  \citealt{CharlotFall00} power-law attenuation law for the younger
  stars). }
\item{we use the M05 SSP library in place of the {\it Padova} one.}
\end{enumerate}
Our results show that changing dust parameters or the modeling of
extinction has only a marginal effect on the shape and normalization
of the number counts. Increasing the fraction of cold gas in the
molecular phase depresses the counts of active objects by $0.3$ dex at
$K \sim 20$, while the passive population is unaffected. A similar
decrease ($0.1$ dex) is observed when using the analytical attenuation
law of \citet{DeLucia07b}; in this case only $\sim49$ per cent of the
original EROs are selected, while an additional $\sim 21$ per cent
(with respect to the original number) is added to the sample. In both
tests on dust attenuation, all the removed objects, as well as the new
inclusions, belong to the active subsample. Since the intrinsic
synthetic SEDs do not change in these tests, the differences in the
colors of model galaxies arise from the different attenuation
laws. Indeed it has been shown by \citet{Fontanot09a} that De Lucia \&
Blaizot's analytical attenuation law is a good approximation of the
mean attenuation law in a {\mor}+{\gs} realization, but the scatter is
large. At the same time, a different choice of dust parameters is able
to change both the mean and scatter of the distribution.

On the other hand, we find that using M05 SSP library gives the
strongest effect on ERO number counts, causing a relevant increase of
EROs at all color cuts; the active population becomes dominant at all
fluxes. In this case the change in the colors of model galaxies is
mainly driven by a change in the intrinsic shape of their synthetic
SEDs. In fig.~\ref{fig:nceros_mar} we show the integral number counts
in this last model: $R-K>5.0$ counts are slightly overpredicted at all
magnitudes, while the $R-K>5.3$ bright counts now agree with
observations; more interestingly the $R-K>6.0$ are now in excellent
agreement with observations, while the $R-K>7.0$ sources are still
under represented in the model. Using a higher molecular fraction or
De Lucia \& Blaizot's analytical attenuation law in conjunction with
the M05 SSP library allows a further (marginal) improvement of model
predictions with data.

\begin{figure}
  \centerline{ \includegraphics[width=9cm]{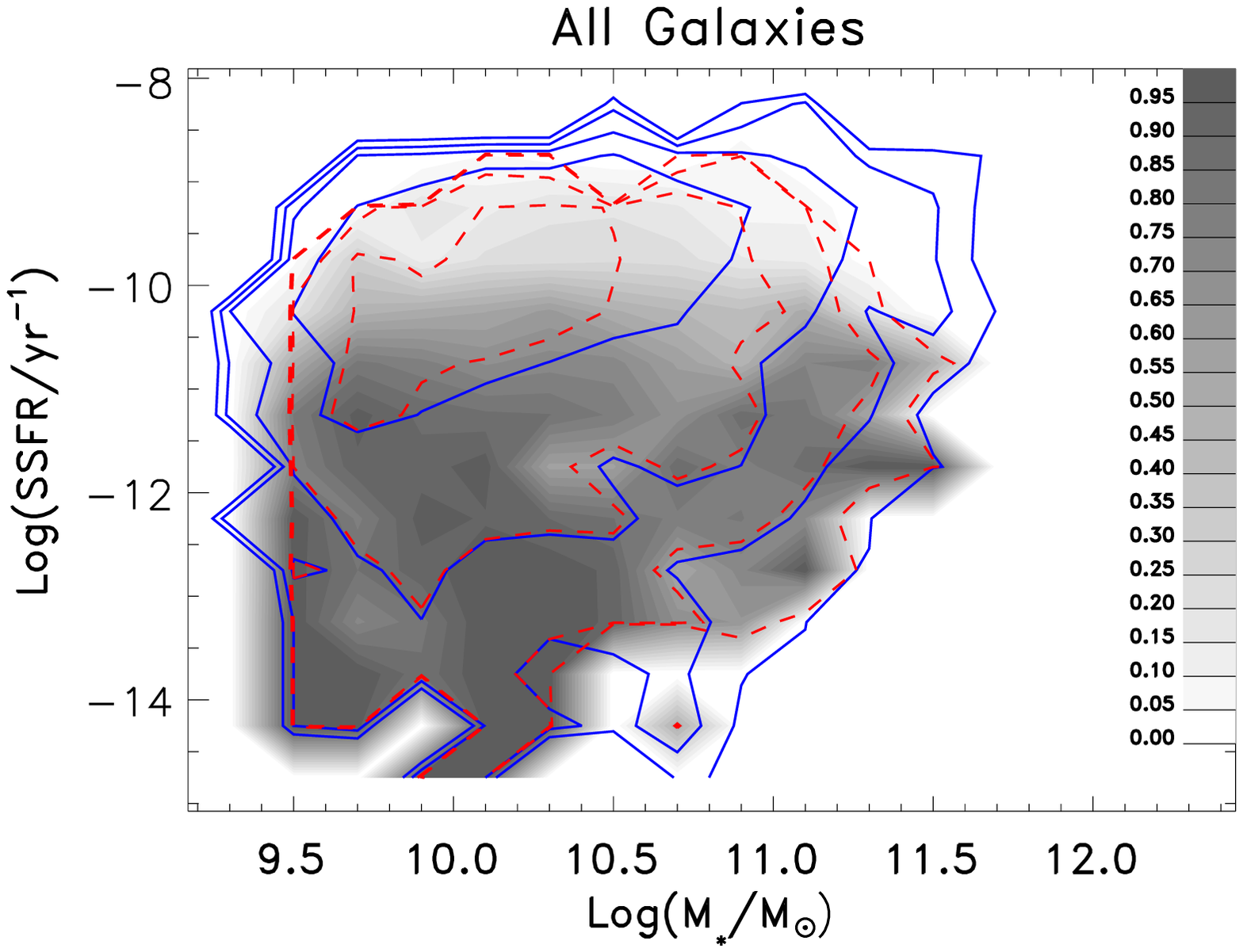} }
  \centerline{ \includegraphics[width=9cm]{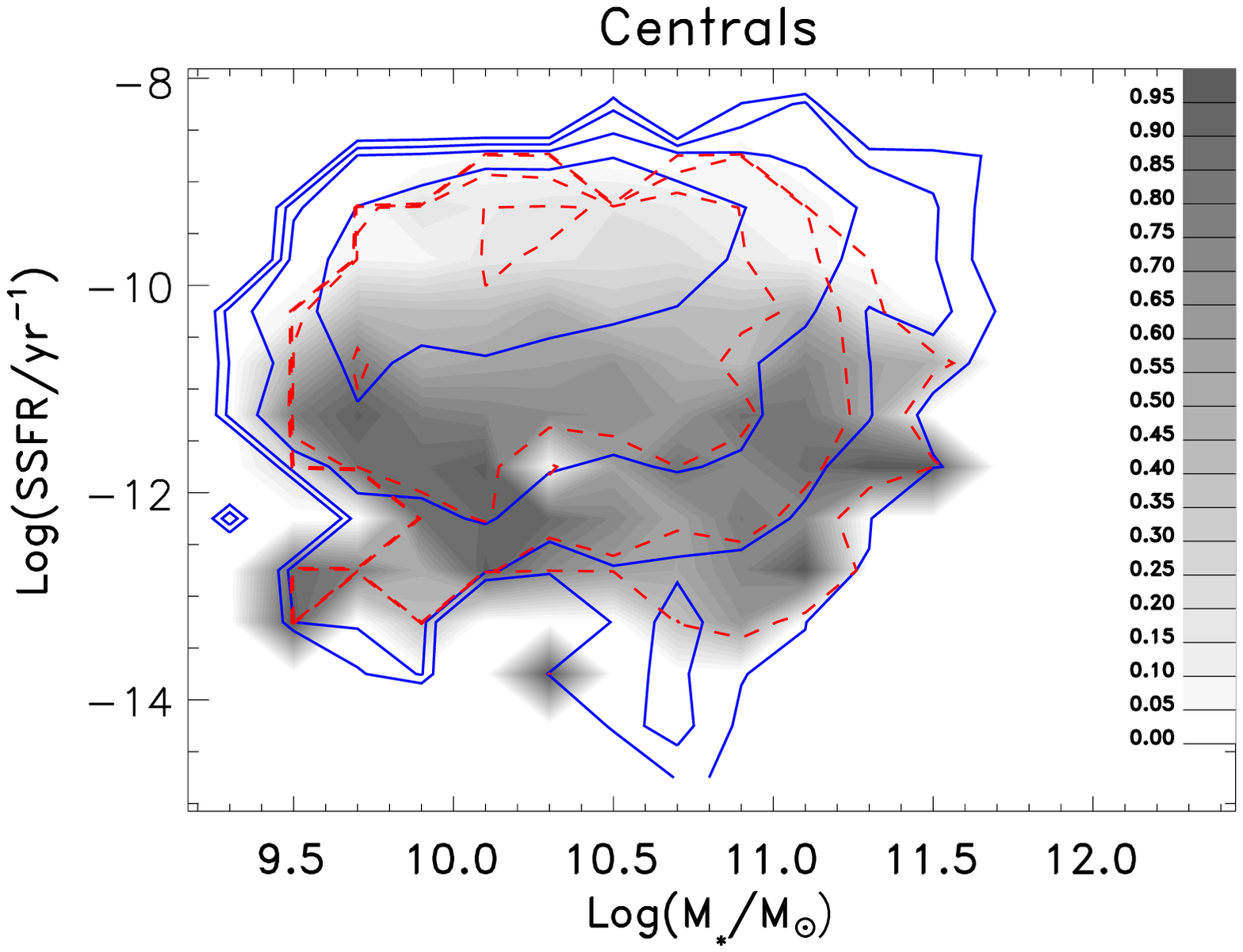} }
  \centerline{ \includegraphics[width=9cm]{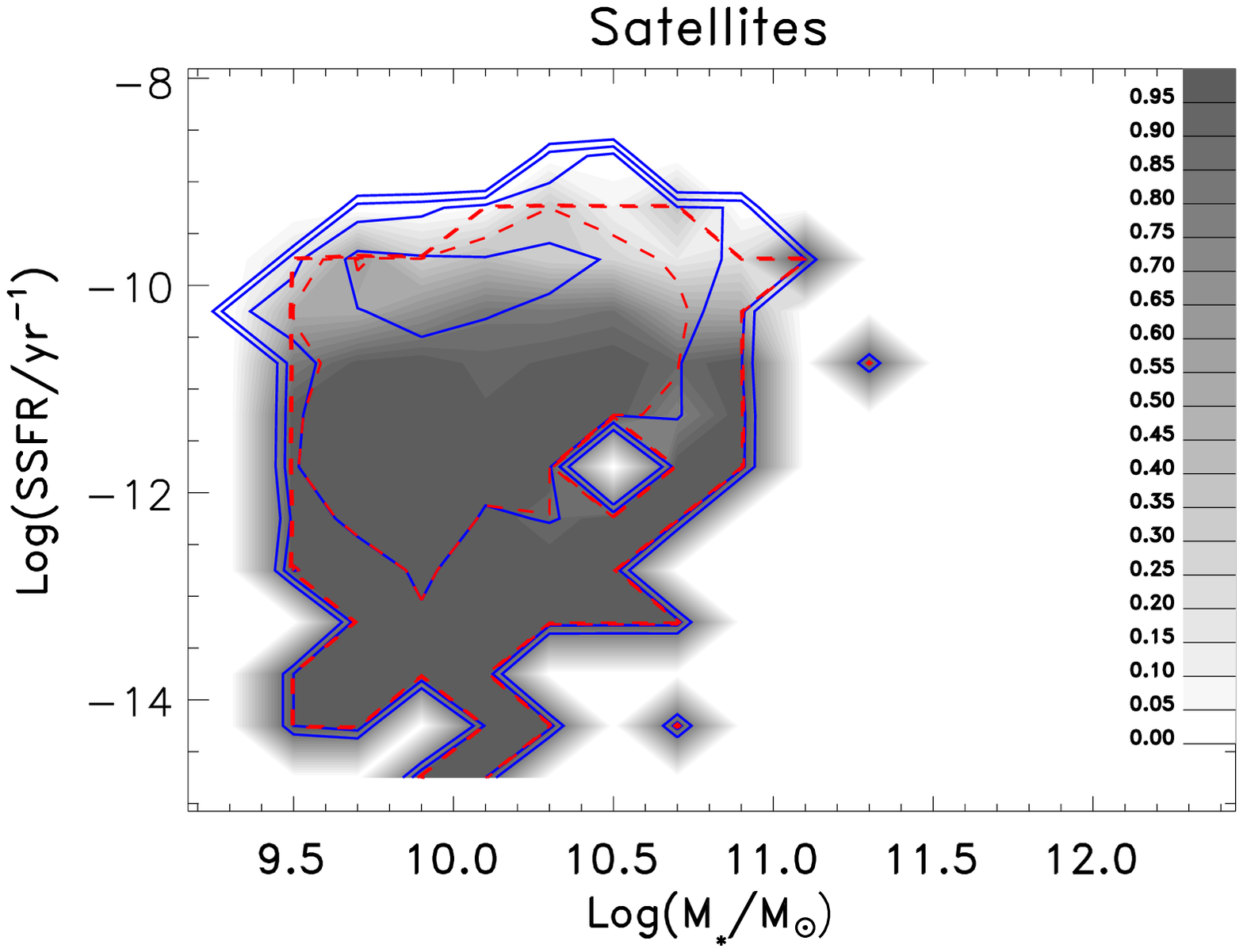} } 
  \caption{Upper panel: 2D probability density of model galaxies in
    the $M_\star$ -- SSFR space.  Blue continuous lines are
    iso-density contours spaced by factors of 10 in density, and are
    relative to the control sample of the all galaxies with $K<22$ and
    in the redshift range $z=[1,2]$.  Dashed red lines give the same
    quantity for the ERO population.  Gray-scale contours give the
    fraction of EROs with respect to the control sample in the same
    $M_\star$--SSFR bin, using the scale given on the right. The mid
    and lower panels report the same quantity for central and
    satellite galaxies respectively.}
  \label{fig:mstar_ssfr}
\end{figure}
\begin{figure}
  \centerline{
    \includegraphics[width=9cm]{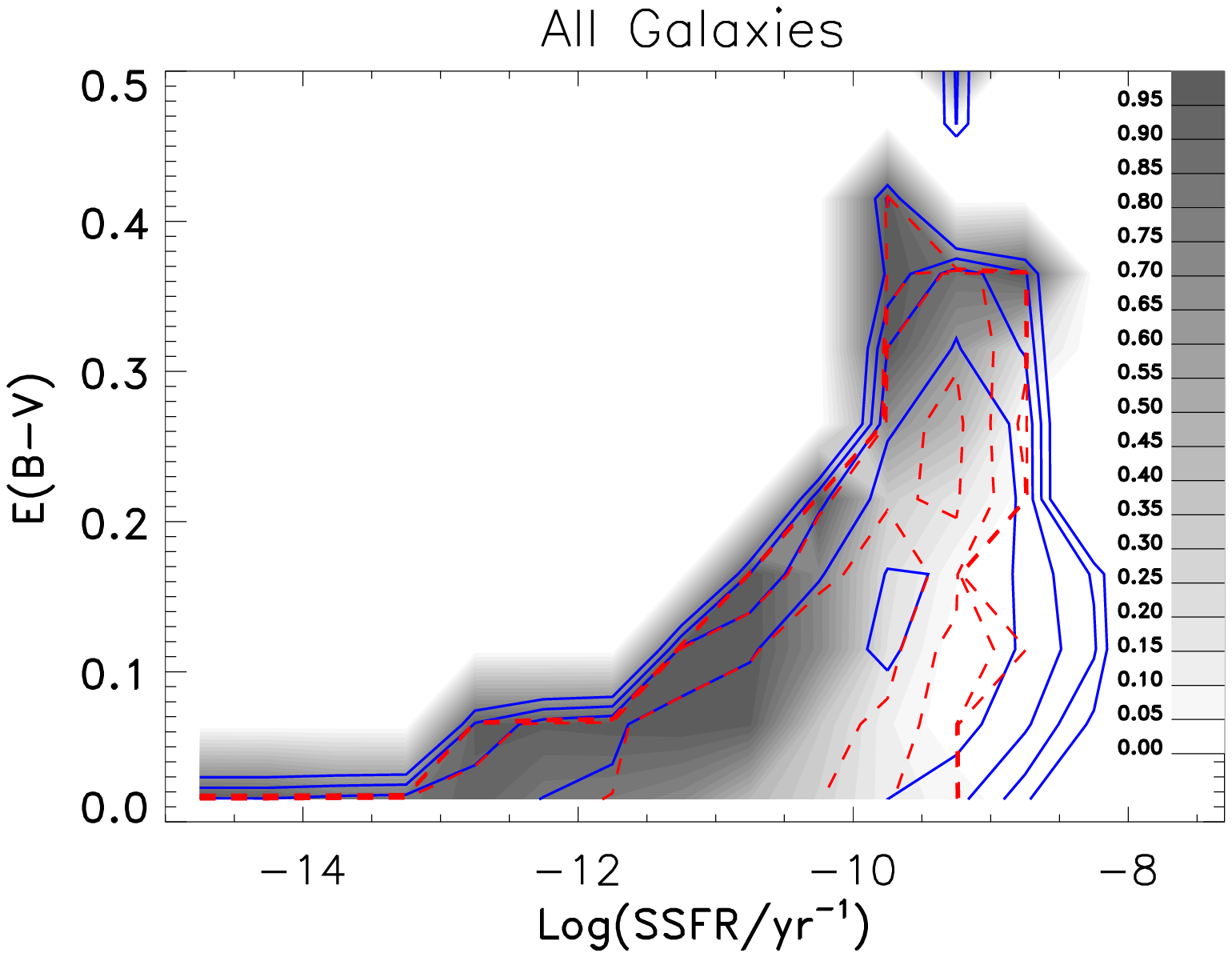} }
  \caption{2D probability density of model galaxies in the SSFR --
    $E(B-V)$ space. Symbols are as in fig.~\ref{fig:mstar_ssfr}.}
  \label{fig:ssfr_ebv}
\end{figure}
\begin{figure}
  \centerline{
    \includegraphics[width=9cm]{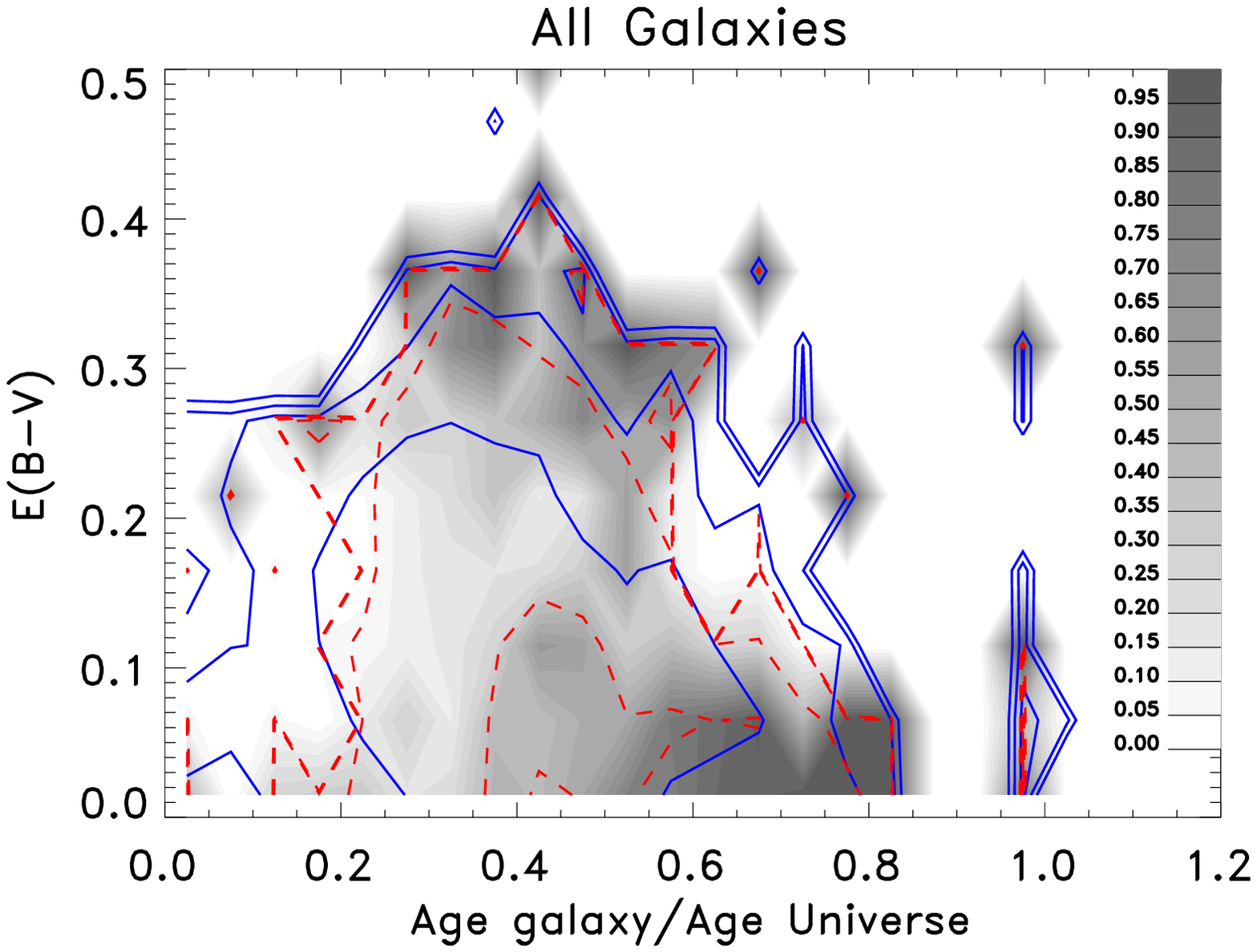} }
  \caption{2D probability density of model galaxies in the space
    defined by mass weighted age, divided by the age of the Universe
    at the redshift of observation, and $E(B-V)$. Symbols are as in
    fig.~\ref{fig:mstar_ssfr}.}
  \label{fig:age_ebv}
\end{figure}
\begin{figure}
  \centerline{ \includegraphics[width=9cm]{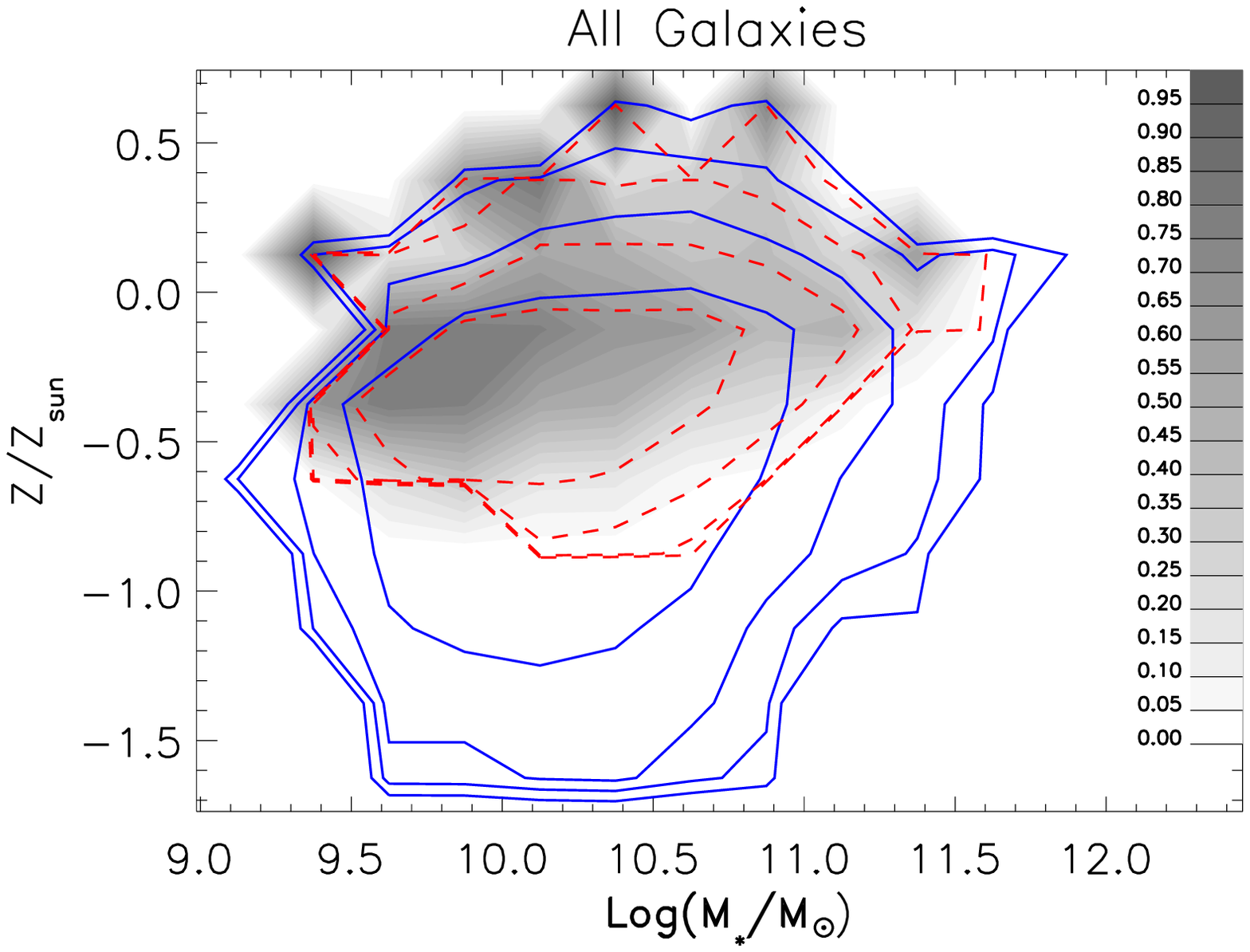} }
  \centerline{ \includegraphics[width=9cm]{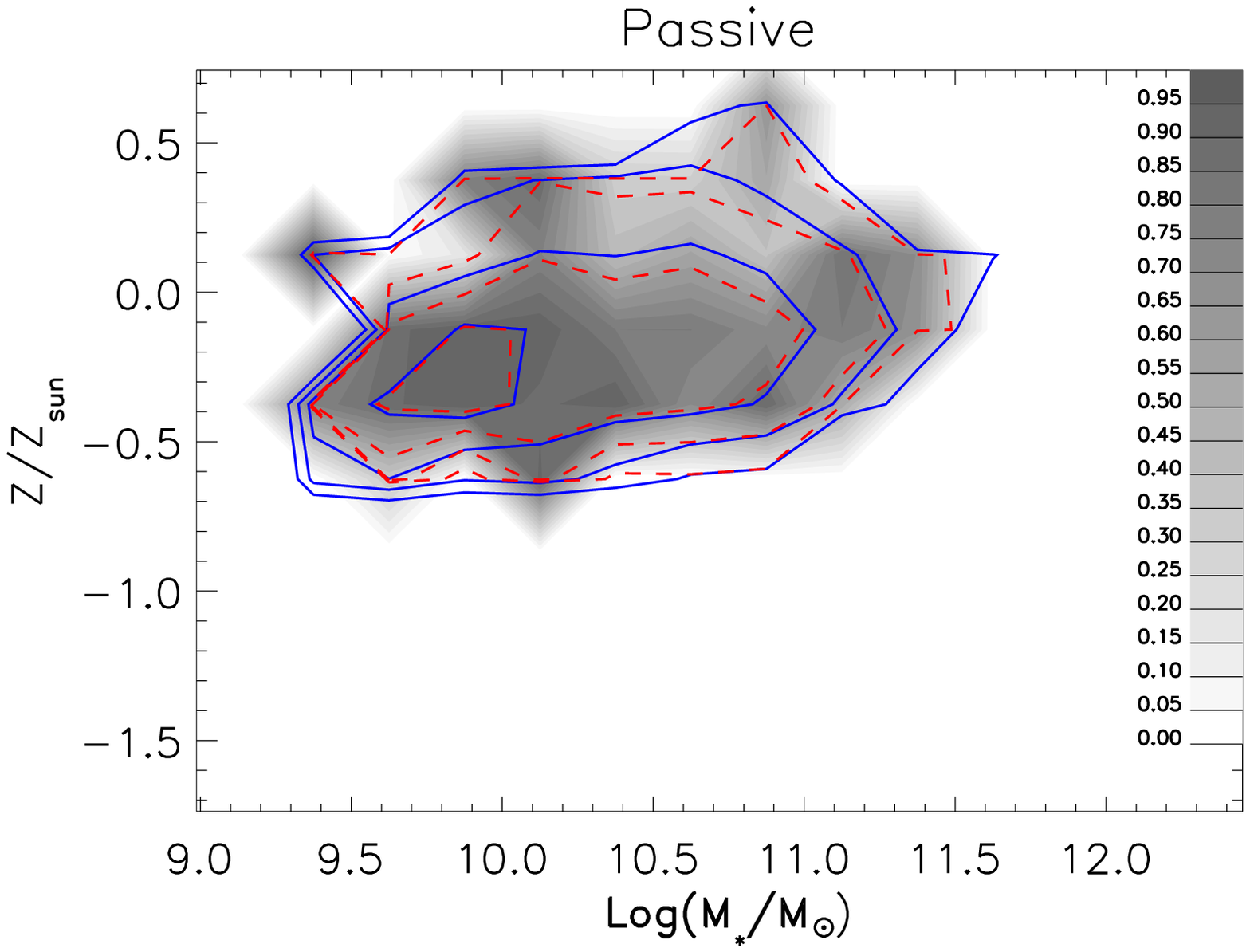} }
  \centerline{ \includegraphics[width=9cm]{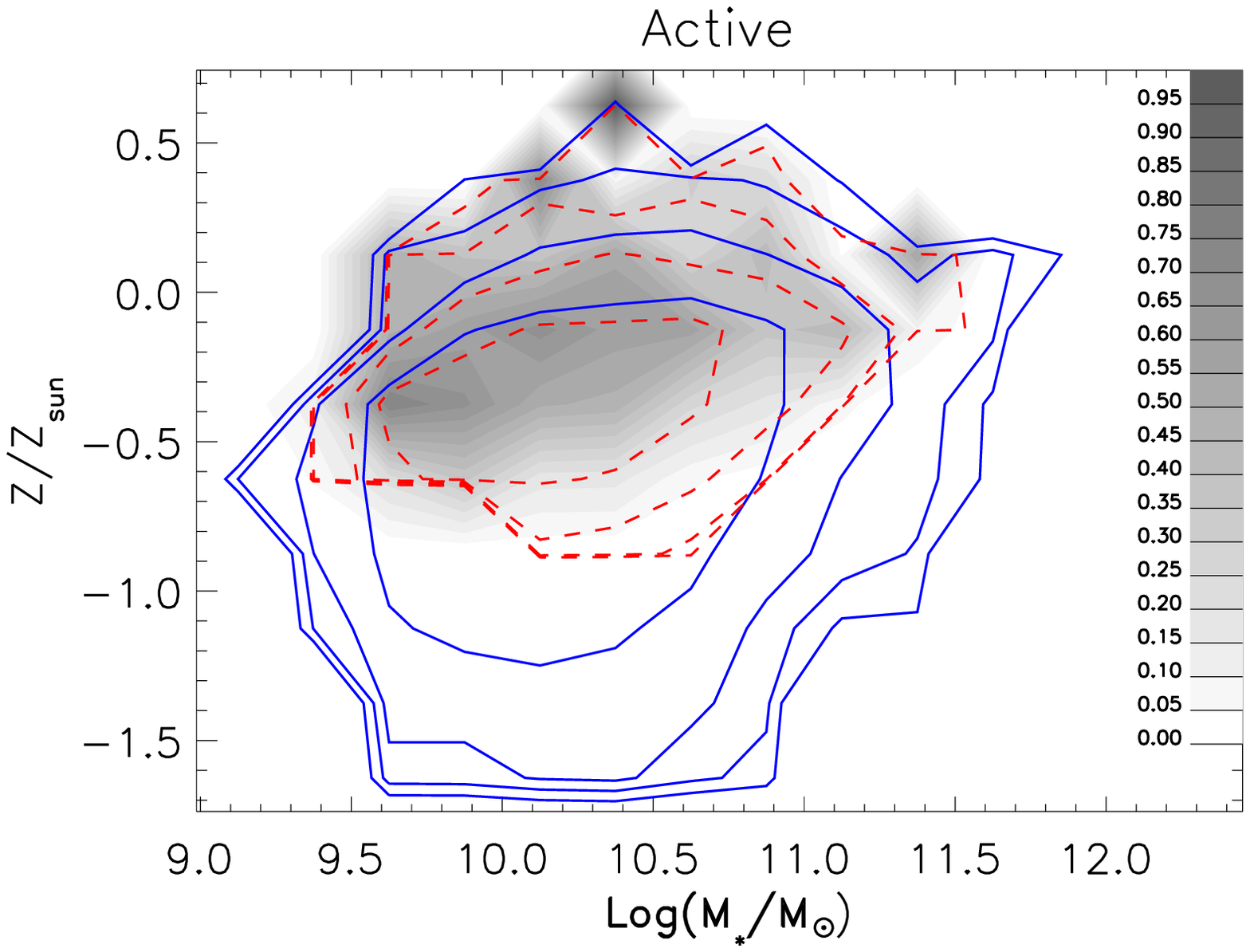} }
  \caption{Upper panel: 2D probability density of model galaxies in
    the $M_\star$ -- $Z_{\star}$ space. Symbols are as in
    fig.~\ref{fig:mstar_ssfr}. The mid and lower panels report the
    same quantity for passive and active galaxies respectively.}
  \label{fig:mstar_Z}
\end{figure}

In fig.~\ref{fig:zd_eros} we compare the redshift distribution of
model EROs with available data. In the left panels we consider the
redshift distribution of the global population \citep{Cimatti02c,
  Miyazaki03, Caputi04} different $K$-band magnitude limits. For our
standard model (solid line), the agreement is good in the probed
range: EROs are mostly present in the redshift range $[0.8,2]$ (mean
redshift $1.34$ at $K<20$), with a sharp low-redshift cutoff which is
well reproduced and a high-redshift tail which is under-represented by
the model. Regarding the normalization, we notice again some excess at
fainter magnitudes and some possible underestimate at $K<19.2$, which
is however not confirmed by other surveys shown in
figure~\ref{fig:nceros}. When the sample is split into active and
passive populations, we see a marginal trend of passive galaxies
peaking at slightly lower redshift, which our model roughly reproduces
(mean redshift $1.27$ and $1.13$ at a depth of $K=20$ for the active
and passive populations respectively); moreover, from the richer
\cite{Miyazaki03} sample we see that the under-reproduction of the
high-redshift tail is mostly due to passive galaxies, while active
EROs are possibly over-represented. The modified models with different
dust parameters or analytical attenuation law predicts redshift
distributions that are almost undistinguishable from the standard
model. The model with the M05 SSPs predicts similar redshift
distribution (mean redshift $1.34$, $1.25$, $1.15$ at $K<20$ for the
total, active and passive populations respectively), but higher number
densities (dashed lines). This is especially relevant for the active
subpopulation. We also show the redshift distribution for the redder
cut $R-K>5.3$ \citep{Simpson06} and we find that they preferentially
select EROs at higher redshift; our model with M05 library reproduces
this trend.

These tests show clearly that, among the three changes in SED modeling
that we tested, the inclusion of TP-AGBs has a strongest effect on the
number counts, and in particular, on the active subpopulation. The
detailed analysis of the effects of this inclusion on the whole galaxy
population and its redshift evolution is clearly beyond the aim of
this paper and we leave it to future work. For this reason, and in
order to allow a direct comparison with the previously published
results for the same model, we decided to discuss the physical
properties of EROs only for the reference model with {\it Padova}
SSPs, which is able to reasonably reproduce counts and redshift
distributions of the $R-K>5.0$ sources, even when we split the
population into active and passive subsamples.  We checked that the
results presented in the rest of this paper do not change
significantly when the model including M05 SSPs is used.

Summing up, the three main points of disagreement of our reference
model with data are: (i) an excess of faint EROs, (ii) an
underestimate of EROs at high color cuts, (iii) an underestimate of
high-redshift passive EROs. These discrepancies are consequences of
the known problems of the model: the excess of small galaxies at
$z\sim1$ that affects most galaxy formation models reflects in the
excess of faint EROs, while because the reddest galaxies are
higher-redshift sources, points (ii) and (iii) are usually related
with the dearth of high-redshift massive galaxies in {\mor}
\citep{Fontanot07b}. It is worth stressing that when TP-AGB stars are
taken into account, the tension at points (ii) and (iii) is
considerably reduced (but not completely solved): this is an hint of
the relevance of the detailed inclusion of this stellar evolutionary
phase in the algorithms for the estimate of stellar masses
\citep{Maraston06}.

\subsection{Properties of the ERO populations.}\label{sec:ero_prop}

The good level of agreement of model EROs with observations at
$R-K>5$, especially in terms of fraction of active and passive
objects, encourages us to use the model to get insight on the
properties of the two ERO populations and on their relation with
sub-mm galaxies and local massive galaxies. To this aim we consider
the predicted physical properties of the $K<22$, $R-K>5$, $1<z<2$ ERO
galaxies, and compare them with a control sample containing all
galaxies with $K<22$ and in the same redshift range. In the following
we will consider only predictions relative to our reference model, but
we checked that the inclusion of any of the modifications we discussed
in the previous section does not affect our conclusions. In
figure~\ref{fig:mstar_ssfr} we show the 2D probability density of ERO
(red dashed lines) and control sources (blue continuous lines) in the
stellar mass -- SSFR space. The gray-scale contours give the fraction
of galaxies in that stellar mass and SSFR bin that meet the selection
criterion. Overall, EROs do not populate a special region in this
space with respect to the global population at similar redshift and,
consistently with \citet{GonzalezPerez09}, they do not show a bimodal
distribution of either passive or active galaxies. Passive galaxies
show a higher probability of satisfying the ERO selection; the most
active galaxies are typically selected out.

A crucial point of galaxy formation models is the different behavior
of central and satellite galaxies; in particular, strangulation of
satellites may artificially boost the number of EROs.  To deepen this
point we separately show the ERO and control distributions of centrals
and satellites in the $M_\star$ -- SSFR space, in the two upper panels
of fig.~\ref{fig:mstar_ssfr}.  As expected, satellites are more
passive than centrals and centrals are more massive than satellites,
and this is true both for the ERO and for the control samples, but the
two populations of active and passive EROs are present in both
sub-samples, and the differences between the distributions in the two
cases are not so strong. We cross-checked this result by comparing
integral number counts of ERO central galaxies with all EROs, finding
only a small difference. In particular, the faint end of the number
counts is reduced by $<0.3$ dex, while the bright end is
unaffected. This is due to the fact that the population of
$M_\star>10^{10}\ \msun$ galaxies is dominated by centrals.  We
conclude that the statistics of model EROs is unaffected by artificial
strangulation of satellites.

\begin{figure}
  \centerline{
    \includegraphics[width=9cm]{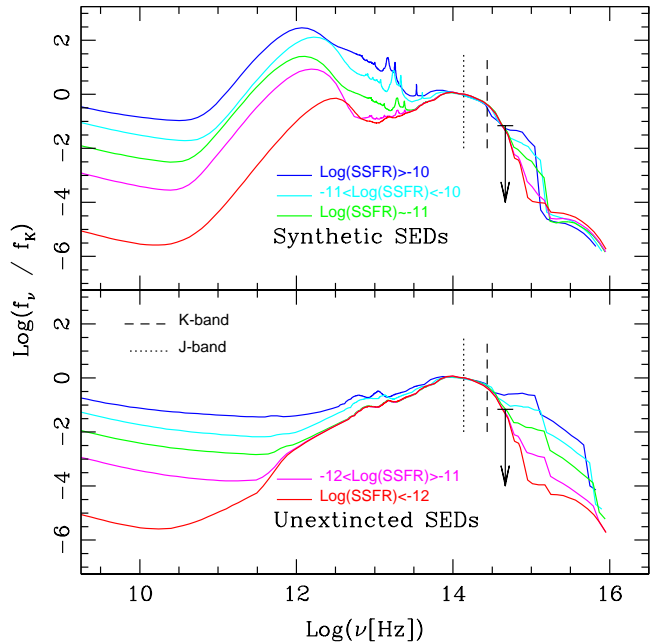} 
  }
  \caption{Comparison between typical SEDs for different activity
    levels. All SEDs have been normalized to the K-band flux. Upper
    panel show the {\gs} synthetic SED, while lower panel show the
    corresponding intrinsic starlight emission. Blue and red lines
    refer to very active (${\rm SSFR} > 10^{-10} yr^{-1}$) and very
    passive (${\rm SSFR} < 10^{-14} yr^{-1}$) objects in the mock
    catalogue. Cyan, green and magenta lines represents typical SEDs
    for EROs with $10^{-10} yr^{-1}< {\rm SSFR} < 10^{-11} yr^{-1}$,
    ${\rm SSFR} \sim 10^{-11} yr^{-1}$, $10^{-11} yr^{-1}< {\rm SSFR}
    < 10^{-12} yr^{-1}$ respectively.}
  \label{fig:seds}
\end{figure}

The absence of a clear bimodality in the distribution of ERO SSFR is
at variance with early ideas on the nature of these objects as either
massive passive galaxies or heavily obscured starbursts.  To further
investigate whether a bimodality is present in the model ERO
population, we define the color excess $E(B-V)$ as:

\begin{equation}
E(B-V) = (B-V)-(B-V)^0
\end{equation}

\noindent
where $(B-V)^0$ represents the intrinsic color computed on the
rest-frame synthetic SEDs before dust extinction, while $(B-V)$ is
computed after dust extinction.  In fig.~\ref{fig:ssfr_ebv} we show
the 2D probability density of control and ERO galaxies in the SSFR --
$E(B-V)$ color space. Contours are as in
figure~\ref{fig:mstar_ssfr}. Also in this space, active and passive
EROs do not populate separate regions, but they define a unimodal
distribution, very similar to that of the overall galaxy
population. Passive galaxies are characterized by low attenuations as
expected, and color excesses range up to a maximum of $\sim0.4$ for
active galaxies. EROs do not populate the region of strongly
star-forming galaxies with $E(B-V)<0.2$. We find a clear bimodality
when we consider the distribution of the fraction of galaxies with
very red colors, which shows two broad peaks for passive galaxies with
$E(B-V)<0.2$ and very active ones, ${\rm SSFR}>10^{-10} yr^{-1}$ and
$E(B-V)>0.2$.  This result does not give support to the assumption of
high attenuations for active EROs made by \citet{Nagamine05}.  In
fig.~\ref{fig:age_ebv} we show the location of control and ERO
galaxies in the space defined by the mass-weighted ages of the stellar
populations, normalized to the age of the Universe at the observation
redshift, and $E(B-V)$.  EROs tend to avoid the region at low age and
attenuation, and again their fraction show a bimodal trend in
$E(B-V)$, but no bimodality is visible in the overall distribution.
Finally, we show in figure~\ref{fig:mstar_Z} the location of EROs in
the stellar mass-metallicity relation. EROs tend to selectively
populate the high metallicity region. This is the result of the
combination of two effects, as can be seen from the smaller panels.
Passive galaxies populate only the high metallicity region, as
expected from evolved stellar populations. The control population of
active galaxies shows a larger spread, but only sources with high
metallicity reach the high levels of obscuration required to be
selected as an ERO.

The distinction between active and passive EROs is an observational
crucial point. From figure~\ref{fig:mstar_ssfr} we see that SSFR
values for EROs have a broad and unimodal distribution, so that moving
the cut around the assumed value of $10^{-11}$ yr$^{-1}$ will not
change much the relative fractions of the two classes. To better
illustrate the continuity of EROs along the SSFR parameter, in
fig.~\ref{fig:seds} we show five typical SEDs (in the observed frame)
for EROs in our mock catalogue with various SSFR values. SEDs are
normalized at the $K$-band wavelength, in order to highlight the
differences in SED shapes. We choose two extreme examples of active
and passive galaxies (${\rm SSFR} > 10^{-10} yr^{-1}$ and ${\rm SSFR}
< 10^{-14} yr^{-1}$) and three objects with intermediate SSFR; in
particular the green line represents an object with SSFR close to the
activity threshold, while the cyan and magenta lines refer to objects
with activity respectively higher and lower than the threshold by less
than a factor of 10.  $K$ and $J$ central wavelengths are shown by
dotted and dashed lines, while at central $R$-band wavelength we
report the upper limit for ERO selection.  This comparison highlights
the continually varying properties of SEDs along the SSFR sequence,
with no clear difference in the predicted spectral properties in the
Near-IR. Larger differences are present at longer wavelengths
($\lambda>10\mu m$), so observations in the Mid-IR and, especially, in
the Far-IR can provide a clear distinction between active and passive
EROs.

\begin{figure}
  \centerline{
    \includegraphics[width=9cm]{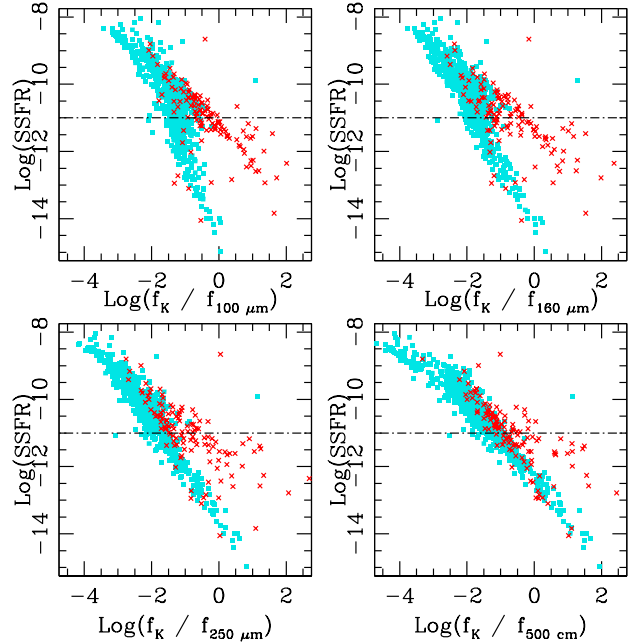} 
  }
  \caption{Predicted relation between the SSFR and the the ratio
    between the $K$-band and the monochromatic fluxes in four {\it
      Herschel} passbands, normalized to the $K$-band flux, for the
    EROs sample. Red crosses and cyan squares correspond to bulge- and
    disc- dominated model galaxies (splitted on the basis of their
    {\it bulge-to-total} ratio, with a threshold value of $0.6$).
    Dot-dashed line represents the assumed threshold between active
    and passive EROs.}
  \label{fig:ssfr_vs_fluxes}
\end{figure}

\begin{figure}
  \centerline{
    \includegraphics[width=9cm]{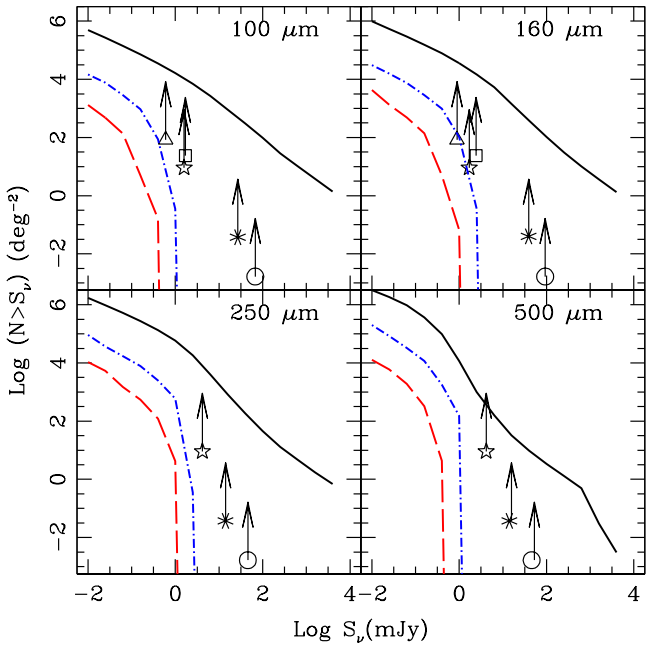} 
  }
  \caption{Predicted number counts in four {\it Herschel}
    passbands. The thick solid, red dashed and blue dot-dashed lines
    refer to the whole population, the passive and active EROs
    respectively. The symbols mark the surface density corresponding
    to the detection of one object in the GOODS-Herschel (triangle),
    PEP GOODS-S (square), HERMES Level-1 (star), Hermes Level-2
    (asterisk) and ATLAS (circle) surveys}
  \label{fig:her_pred}
\end{figure}

Such observations can in principle be carried out by the {\it
  Herschel} Far-IR space telescope.  We show in
fig.~\ref{fig:ssfr_vs_fluxes} the predicted relation between the SSFR
of EROs and the ratio between $K$-band and monochromatic fluxes in
four bandpasses covered by the instruments on board {\it Herschel}, in
our smaller mock catalogue where IR fluxes are computed using full
{\gs} calculations. All four colors show a clear correlation with
SSFR; we find much weaker trends for fluxes at shorter wavelengths, as
those currently probed by the {\it Spitzer} space telescope. These
correlations are stronger at high levels of specific star formation,
while passive galaxies split into two distinct sequences. This
splitting relates to the modeling of star formation in bulges or
discs: the upper sequence is dominated by galaxies with {\it
  bulge-to-total} mass ratios $>0.6$ \citep[as in][]{Fontanot09a},
denoted by red crosses while cyan squares correspond to disc-dominated
galaxies. We checked that this splitting is mainly due to the presence
of a population of passive and high-metallicity discs, that give rise
to the lower branch. Getting back to active galaxies, it is evident
that a Far-IR follow-up of the already-known EROs would be of great
importance in order to have a better understanding of the properties
of these objects, and in particular on their star formation rate. We
present in fig.~\ref{fig:her_pred} the predicted integrated number
counts for passive and active EROs (red dashed and blue dot-dashed
lines respectively). In each panel, we also mark the surface densities
corresponding to one detection in the {\it Herschel} Key
Programmes\footnote{http://herschel.esac.esa.int/} (GOODS-Herschel,
HERMES, ATLAS, see caption for symbol coding).  The result is not
exciting: we predict that it is unlikely that these surveys will
detect a large number of EROs.  Incidentally, in the same figure we
report our prediction for the integrated number counts of the whole
galaxy population (thick solid line). Our differential number counts
are in good agreement (difference below $0.2$ dex) with the results of
\citet{Lacey09} for infrared fluxes $S > 1$ mJy; at fainter fluxes
      {\mor} predicts a higher abundance of sources. It will be soon
      possible to compare this prediction with the forthcoming {\it
        Herschel} observations.

\subsection{Evolutionary sequences in \mor.}
\label{sec:sequence}

In this section we focus on the relation between EROs (defined as
objects with $R-K>5.0$ and $K<22$), sub-mm galaxies (defined as
galaxies with a predicted $850\ \mu$m flux $f_{850 \mu m}>0.5$ mJy)
and local massive galaxies ($M_\ast>10^{11}M_\odot$ at $z=0$).

In order to use IR templates in place of the more expensive {\gs}, we
define sub-mm bright galaxies as those objects with $f_{850 \mu m} >
0.5$ mJy, a flux where number counts based on the two methods are very
similar (sec.~\ref{sec:mock}) and only slightly higher that the limit
reached in deep fields with currently available sub-mm telescopes.

For these three categories we follow their evolution in the redshift
range $0.5<z<4.5$, with a time sampling of 100 Myr; for simplicity we
restrict to the main progenitor branch of the galaxy merger tree.  We
check to what extent and in what time order a galaxy belonging to one
category is observed as a member of another category.

\begin{figure*}
  \centerline{
    \includegraphics[width=15cm]{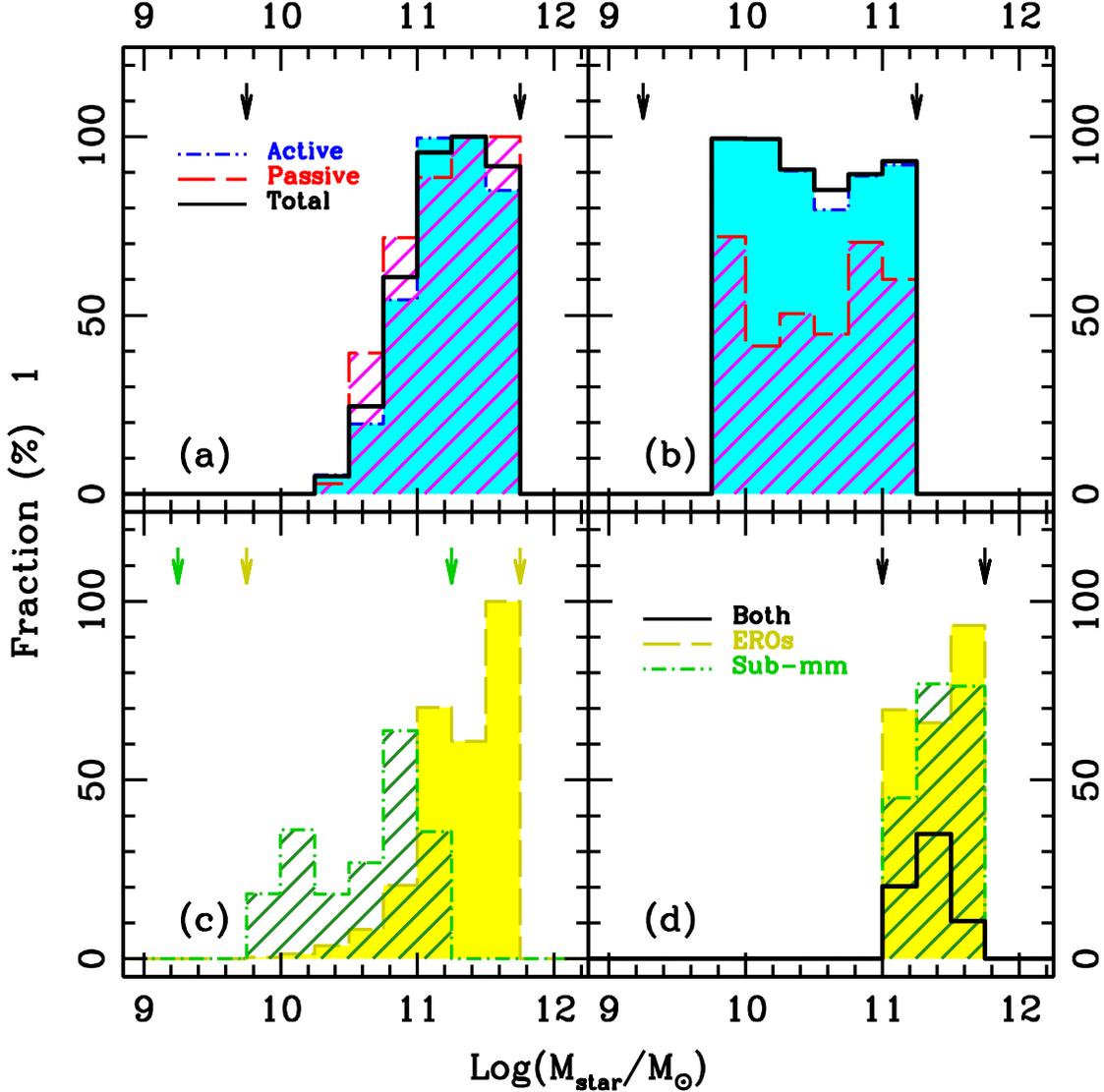} 
    }
  \caption{{\it Panel (a)}: Fraction of EROs with a sub-mm main
    progenitor with $f_{850 \mu m}>0.5$ mJy in {\mor}. Thick solid,
    red dashed (diagonal texture) and blue dot-dashed (cyan shading)
    histograms refer to the total and passive and active EROs
    respectively. {\it Panel (b)}: Fraction of sub-mm galaxies that
    are the main progenitor of an ERO in {\mor}. Thick solid histogram
    refers to the total number of sub-mm galaxies, while the red
    dashed (diagonal texture) and blue dot-dashed (cyan shading)
    histograms indicate the properties of the descendant as indicated
    in the caption. {\it Panel (c)}: Fraction of EROs (solid histogram
    and yellow shading) and sub-mm galaxies (green dot-dashed
    histogram and diagonal texture) that are main progenitors of a
    local massive galaxy ($M_\ast > 10^{11} M_\odot$) as a function of
    their stellar mass. {\it Panel (d)}: Fraction of massive ($M_\ast
    > 10^{11} M_\odot$) galaxies at $z<0.2$ with a EROs (solid
    histogram yellow shaded area), a sub-mm progenitor ($f_{850 \mu
      m}>0.5$ mJy green dot-dashed histogram and diagonal texture) or
    both (thick solid histogram). In all panels the vertical arrows
    indicate the mass limits for the reference sample of EROs, sub-mm
    and massive galaxies.}
  \label{fig:eros_frac}
\end{figure*}

As a first figure, the fraction of EROs that are seen {\it at the same
  time} as sub-mm sources is low, 0.1 per cent, while the fraction of
sub-mm galaxies that are seen as (active) EROs is 5 per cent; then,
the overlap between the two categories is rather small. In
fig.~\ref{fig:eros_frac} we show some diagnostics of the evolutionary
path of galaxies in some specific category as a function of their
galaxy stellar mass. The vertical arrows mark the mass limits within
which the diagnostic has sufficient statistics (only mass bins with at
least 10 objects are considered). Panel (a) of
Fig.~\ref{fig:eros_frac} shows the fraction of EROs that are
classified as sub-mm galaxy at an earlier epoch as a function of their
stellar mass at the ERO epoch; the three histograms give this quantity
for active, passive and all EROs (as specified in the legend). Almost
all EROs with stellar masses $\ga 10^{11}\ \msun$ have had a sub-mm
ancestor as a main progenitor, irrespective of their level of
activity. The connection is soon lost at lower masses.  In panel (b)
of fig.~\ref{fig:eros_frac} we select $z>1$ sub-mm sources and check
whether they are predicted to be the main progenitors of a (passive or
active or either) ERO. Here the stellar mass refers to the sub-mm
selected galaxies. The resulting fraction is high and does not depend
on stellar mass for $M_\star > 10^{9.75} M_\odot$. This means that
most sub-mm galaxies have an ERO descendant, which is in most cases an
active ERO, while only roughly half of them have a passive ERO
descendant.

We then consider the expected relation of the two high-redshift
categories with local massive galaxies.  Panel (c) of
figure~\ref{fig:eros_frac} shows the fraction of EROs and sub-mm
galaxies that are {\it main} progenitors of a $z=0$ massive galaxy, as
a function of their stellar mass at the observation redshift.  More
massive EROs and sub-mm galaxies are more likely to be main
progenitors, and the relation is tighter for EROs, while sub-mm
galaxies have a relatively modest, $\sim30$ per cent, probability.  A
different result is obtained when the main progenitors of local
massive galaxies are considered, as in panel (d) of
figure~\ref{fig:eros_frac}: here we report the fraction of local
massive galaxies that have an ERO or sub-mm main progenitor, as a
function of $z=0$ stellar mass (obviously larger than
$10^{11}\ \msun$).  The main progenitor of a massive galaxy used to be
an ERO more than 80 per cent of times and a sub-mm galaxy $\sim70$ per
cent of times; as a matter of fact, each object may be seen as an ERO
or sub-mm galaxy at many redshifts. So, the low probabilities shown in
panel (c) imply that many massive EROs or sub-mm galaxies end in
massive galaxies not as main progenitors. In other words, consistently
with the hierarchical clustering scenario, many massive galaxies will
have not one but several ERO and/or sub-mm progenitors. However, panel
(d) shows also that only $\sim20$ per cent of the main progenitors of
local massive galaxies were both sub-mm galaxies and EROs (in this
order) in their past.

The results in fig.~\ref{fig:eros_frac} clearly depend on the adopted
flux limit of sub-mm bright galaxies.  Lowering (or raising) this flux
has the effect of decreasing (increasing) the mass cutoff in panel
(a), thus increasing (decreasing) the fraction of all EROs with sub-mm
progenitors at intermediate masses. In panel (d), the fractions of
sub-mm and ERO+sub-mm main progenitor increase (decrease) accordingly.
Instead, the fractions shown in panels (b) and (c) are hardly affected
as long as the flux limit is below 2 mJy.

\begin{figure}
  \centerline{
    \includegraphics[width=9cm]{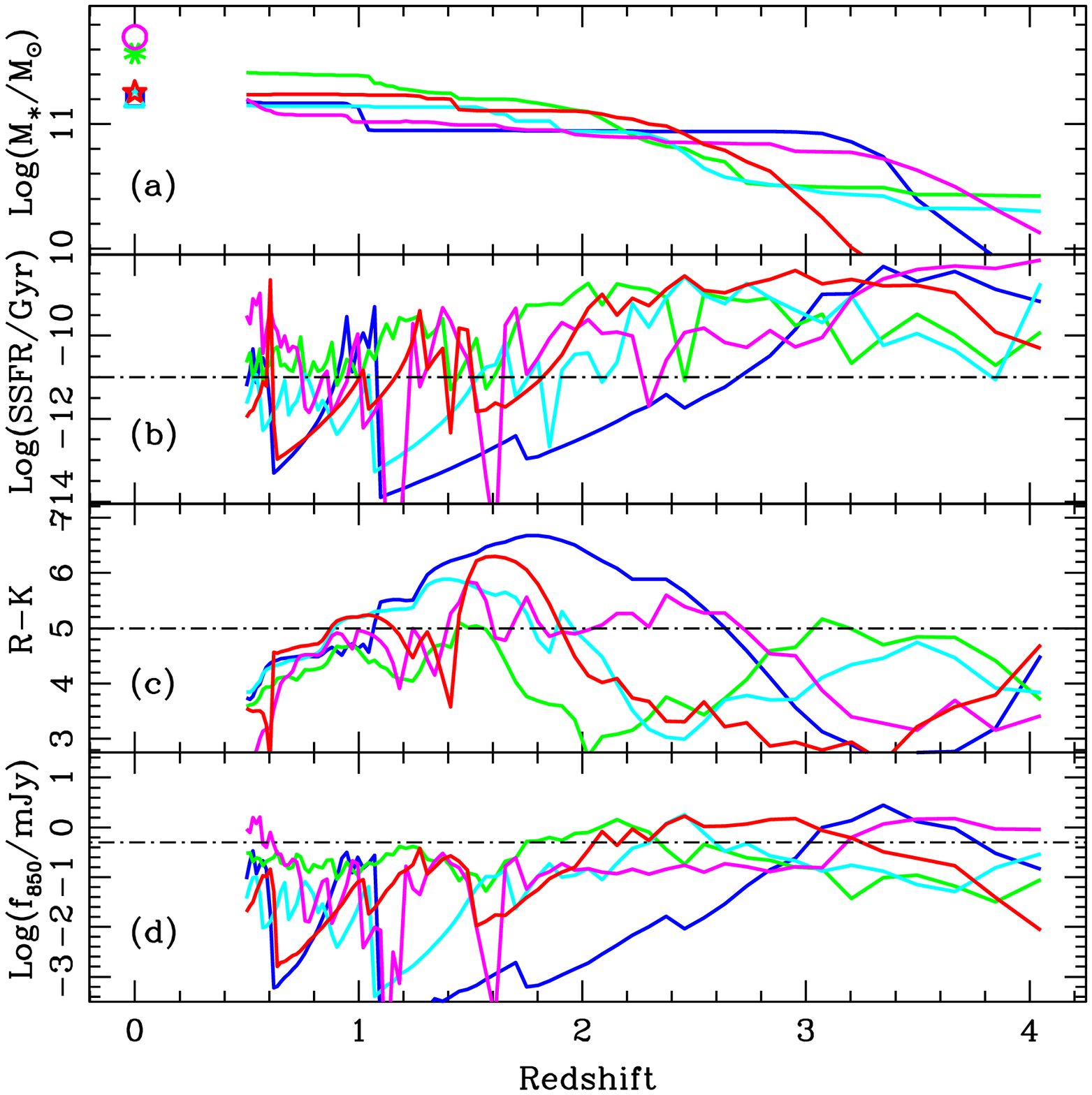} 
  }
  \caption{The redshift evolution of the total mass ({\it panel a}),
    SSFR ({\it panel b}), observer frame R-K color ({\it Panel c}),
    and observer frame $850 \mu$m flux ({\it Panel d}) for 5
    representative galaxies. In each panel the same greyscale
    corresponds to the same object. Empty symbols correspond to the
    $z=0$ mass of the 5 galaxies.}
  \label{fig:flux_evo}
\end{figure}

In fig.~\ref{fig:flux_evo} we consider the evolution of 5
representative model objects, chosen between the $\sim25$ per cent
explicitly following the sub-mm -- EROS -- massive galaxy evolutionary
sequence. In the four panels, from top to bottom, we show the total
stellar mass (symbols mark the $z=0$ mass), the specific star
formation rate, the $R-K$ color and the $850 \mu$m flux.  The
diversity of evolutionary tracks is evident from the figure, as well
as the complexity of star formation histories.  In most cases the two
phases are well separated along the redshift evolution of the
object. Some model galaxy undergoes repeated independent ERO (and
sub-mm) episodes along their evolution; some of them can be seen both
as an active and a passive ERO at different redshifts. Even if in most
cases the ERO stage follows the sub-mm bright period, it is also
possible that, as a result of late star formation activity, a model
galaxy may be recognize as a sub-mm bright galaxy, even if it was a
passive ERO at previous cosmic times. Interestingly, active EROs tend
to have low sub-mm fluxes, consistently with the low fraction of
galaxies that are at the same time ERO and sub-mm.  These results show
that the galaxy evolution sequences in our galaxy formation model
based are more complex compared to the simple one obtained by
\cite{Granato04}, that predict that massive galaxies go through
sub-mm, quasar, ERO and passive phases.

\section{Conclusions}\label{sec:final}

In this paper we have discussed the statistical properties of the ERO
population as predicted by the {\mor} galaxy formation model, in terms
of number counts, redshift distributions and active fractions of
$R-K>5$ sources. We generated mock photometric catalogues by
interfacing {\mor} with the spectro-photometric + radiative transfer
code {\gs}, and created large samples of model galaxies (so as to have
good statistics on the rare EROs). We used both the {\it Padova} and
the M05 SSP libraries; the latters include the contribution of TP-AGB
stars. We combined the resulting UV-to-Near-IR synthetic SEDs,
computed without the expensive Far-IR emission, with the
\cite{Rieke09} IR template library to estimate IR fluxes longward $3
\mu m$.  We then selected samples of EROs (and sub-mm galaxies) by
applying the corresponding color and flux cuts.

We found that our standard model with {\it Padova} SSP library is able
to reproduce the overall number counts and redshift distribution of
the $R-K>5$ sources; the agreement worsens when redder cuts are
considered. Other discrepancies, like the excess of faint EROs at deep
$K$ fluxes and the underestimate of the number of distant passive
EROs, are consequences of well-known points of tension of the model
with data, like the dearth of massive $z>2$ galaxies
\citep{Fontanot07b,Cirasuolo08,Marchesini09}, or the excess of
$z\sim1$ small galaxies \citep{Fontanot09b}; this analysis does not
reveal new discrepancies.  Interestingly, {\mor} is able to roughly
reproduce the substantial contribution (of the order of $\sim 50\%$,
see e.g. \citealt{Cimatti02a,Miyazaki03}) of active galaxies to the
global ERO population; here we define active galaxies as those with
${\rm SSFR} > 10^{-11}$ Gyr. We ascribe this success to the higher
SSFR predicted by the model at higher redshifts
\citep{Fontanot07b,Santini09}.

We showed that the ERO number counts increase considerably when using
the M05 SSP library.  The addition of TB-AGBs gives a strong boost to
the restframe SEDs of active galaxies longward $6000$ \AA, reddening
them and thus increasing the number of active EROs.  This leads to a
good level of agreement even at very red cuts ($R-K>6$), though it
does not completely solve the lack of very red passive galaxies at
higher redshift. This highlights the importance of an accurate
modeling of galactic synthetic SEDs when comparing the prediction of
theoretical models of galaxy formation to observations, and confirms
that TP-AGBs are a relevant ingredient for the correct reproduction of
the high-redshift Universe \citep{Maraston06}. The details of the
modeling of dust attenuation also play a role: we checked that both
increasing the fraction of cold gas locked into molecular clouds (as
suggested by \citealt{LoFaro09} to reproduce the properties of Lyman
break Galaxies at $z \ga 4$) or using the analytical approach proposed
by \citet{DeLucia07b} based on an universal attenuation law slightly
reduces the number of active EROs ($\lesssim 0.3$ dex at $K \sim 20$).

We then considered the physical properties of the predicted ERO
population with respect to a control sample of non color-selected
galaxies with $K<22$ and $1<z<2$.  Overall the physical properties of
EROs do not differ significantly from those of a typical galaxy in the
same redshift range, apart that active galaxies with low dust
attenuation and low metallicities are excluded.  We do not see any
signature for a significant bimodality in the ERO population: the
probability of a galaxy to be selected peaks in two regions of the
SSFR--$E(B-V)$ space, corresponding to moderately attenuated passive
galaxies and strongly attenuated ($E(B-V)>0.2$) active galaxies, but
this trend is not visible in the overall ERO distribution.  This
result is in agreement with \citet{GonzalezPerez09}. Deep fields in
the Far-IR, from 100 to 500 $\mu$m, are the best way to constrain the
SSFR of these objects, but presently planned surveys with {\it
  Herschel} will detect a small number of EROs at best.

We also tested the possible relation between EROs and both the sub-mm
($f_{850 \mu m}>0.5$ mJy) galaxies and the local massive ($M_\ast >
10^{11}$ M$_\odot$) galaxies. This was done to test the suggestion
that these different populations reflect well-defined stages of a
typical evolutionary sequence of massive galaxies, like sub-mm
galaxies evolving into EROs and later into massive passive galaxies.
We found that most massive EROs have a sub-mm bright main progenitor
and evolve into a massive galaxy at $z=0$, and that many $z=0$ massive
galaxies have an ERO or a sub-mm main progenitor, but only a minor
fraction, $\sim25$ per cent, follow the sub-mm -- ERO -- massive
galaxy sequence.  Moreover, massive galaxies may easily have more than
one sub-mm and/or ERO progenitor.  In fact, star formation histories
predicted by {\mor} are far more complex and exhibit a great degree of
diversity, with only a small sample following the well defined
sequence described above.

Our work shows that EROs can be used as a powerful constraint for
theories of galaxy formation and evolution, but a better and deeper
understanding of the distribution of EROs in a configuration space
defined by the physical quantities, like star formation activity,
stellar mass, metallicity, stellar population ages and dust
obscuration, is needed.  Unfortunately, planned surveys with {\it
  Herschel} may not reach the combination of depth and area required
for detecting and analysing statistically significant samples of EROs.
On the other hand, the same instruments may provide the required
spectroscopic and photometric follow-up in order to better understand
the real nature of already detected EROs.

\section*{Acknowledgments}
We thank Laura Silva for making the {\gs} code available and for many
discussion; Claudia Maraston, Chiara Tonini and Bruno Henriquez for
their help in including the M05 SSP library in our algorithm and for
stimulant discussions; Gianluigi Granato, Lucia Pozzetti for useful
comments. Some of the calculations were carried out on the PIA cluster
of the Max-Planck-Institute f\"ur Astronomie at the Rechenzentrum
Garching.

\bibliographystyle{mn2e}
\bibliography{fontanot}

\label{lastpage}

\end{document}